\documentclass[a4paper,twocolumn,11pt,accepted=2024-09-03]{quantumarticle}
\pdfoutput=1
\usepackage[utf8]{inputenc}
\usepackage[english]{babel}
\usepackage[T1]{fontenc}
\usepackage{amsmath,dsfont}
\usepackage{hyperref}
\usepackage{bm}
\usepackage[numbers,sort&compress]{natbib}

\usepackage{graphicx}
\usepackage{amssymb,amsmath}
\usepackage{amsthm}
\usepackage{bm}
\usepackage{dcolumn}
\usepackage{subfigure}
\usepackage{float}
\usepackage[OT1]{fontenc} 
\usepackage{url}
\usepackage{slashed}
\usepackage{color}
\usepackage{verbatim}
\usepackage{txfonts}
\usepackage{soul}
\usepackage{subfigure}
\usepackage{comment}
\usepackage{amssymb}
\usepackage{physics}
\usepackage{enumitem}

\usepackage{tikz}
\usepackage{lipsum}

\def\bea{\begin{eqnarray}}
\def\eea{\end{eqnarray}}

\def\ba{\begin{array}}
\def\ea{\end{array}}

\def\Tr{\text{Tr}}

\usepackage{environ}

\NewEnviron{myequation}{%
    \begin{equation}
    \scalebox{0.87}{$\displaystyle{\BODY}$}
    \end{equation}
    }

\begin{document}

	\title{Efficient Classical Shadow Tomography through Many-body Localization Dynamics}
	
	\author{Tian-Gang Zhou}
    \affiliation{Institute for Advanced Study, Tsinghua University, Beijing, 100084, China}

	\author{Pengfei Zhang}
	\email{pengfeizhang.physics@gmail.com}
	\affiliation{Department of Physics, Fudan University, Shanghai, 200438, China}
        \affiliation{Shanghai Qi Zhi Institute, AI Tower, Xuhui District, Shanghai 200232, China}

\maketitle

\begin{abstract}
Classical shadow tomography serves as a potent tool for extracting numerous properties from quantum many-body systems with minimal measurements. Nevertheless, prevailing methods yielding optimal performance for few-body operators necessitate the application of random two-qubit gates, a task that can prove challenging on specific quantum simulators such as ultracold atomic gases. In this work, we introduce an alternative approach founded on the dynamics of many-body localization, a phenomenon extensively demonstrated in optical lattices. Through an exploration of the shadow norm---both analytically, employing a phenomenological model, and numerically, utilizing the TEBD algorithm---we demonstrate that our scheme achieves remarkable efficiency comparable to shallow circuits or measurement-induced criticality, resulting in a significant improvement in the exponential exponent compared to the previous classical shadow protocol.
 Our findings are corroborated through direct numerical simulations encompassing the entire sampling and reconstruction processes. Consequently, our results present a compelling methodology for analyzing the output states of quantum simulators.
\end{abstract}
\section{Introduction}
Rapid developments in quantum science and technology have ushered in an era where quantum devices are beginning to demonstrate their supremacy \cite{Arute:2019zxq,PRXQuantum.2.017003,PhysRevLett.127.180502,PhysRevLett.127.180501,Mi:2021gdf,Ebadi:2020ldi,Semeghini:2021wls,2022SciBu..67..240Z}. However, the efficient storage and analysis of the output quantum states from these devices have emerged as central challenges. A complete description of $N$ qubits requires $4^N-1$ parameters, which in turn necessitates an exponential number of non-local measurements \cite{2012NJPh...14i5022F,2015arXiv150801907O,2015arXiv150801797H}. Fortunately, we are usually interested in physical properties that do not require a complete understanding of the density matrix, such as physical observables or fidelity. Under these circumstances, \textit{classical shadow tomography} can be employed as an efficient method to extract multiple physical properties with a small number of measurements \cite{2017arXiv171101053A,2020NatPh..16.1050H,2019arXiv191010543P,PRXQuantum.2.030348,2021arXiv210505992A,PRXQuantum.2.010307,2021arXiv211002965L,PhysRevLett.127.110504,BabbushMatchgateShadowsFermionic2023,2021arXiv210612627H,PhysRevResearch.4.013054,PhysRevResearch.5.023027,2023Quant...7.1026A,jaffeClassicalShadowsPauliinvariant2024,PhysRevLett.130.230403,2022arXiv220912924B,2022arXiv221109835A,2023arXiv230715011I,2023arXiv230801653A,Koh2022classicalshadows,TongLearningConservationLaws2023}.

When conducting measurements on a quantum system, the first step involves specifying the measurement basis. Instead of carrying out repeated measurements in a predetermined basis, classical shadow tomography employs randomized measurements \cite{Qi:2019rpi,PhysRevA.99.052323,2019Sci...364..260B,2023NatRP...5....9E}. This technique involves selecting an independent random unitary operation, denoted as $U$, from the ensemble $\mathcal{E}_U$. This operation is applied before measurements are performed on a computational basis. Denoting the post-measurement state as $|b\rangle$, we obtain a classical snapshot $\sigma_{U,b}=U^\dagger|b\rangle \langle b| U$ of the original density matrix $\rho$ with a probability $p_{U,b}=\text{tr}(\sigma_{U,b}\rho)$. The resulting average snapshot $\sigma$ is related to $\rho$ through a quantum channel $\mathcal{M}$ \cite{2020NatPh..16.1050H}: 
\begin{equation}\label{eq:channel}
\sigma= \sum_{b}\mathds{E}_{\mathcal{E}_U}[\sigma_{U,b}~p_{U,b}]\equiv \mathcal{M}(\rho).
\end{equation}
Hence, physical observables can be represented as $\langle O \rangle = \text{tr}(O \mathcal{M}^{-1}(\sigma)) = \text{tr}(\mathcal{M}^{-1}(O) \sigma)$. The computation of $\mathcal{M}$ and the subsequent application of its inverse occur on a classical computer. For the estimation of Eq.~\eqref{eq:channel}, it's essential that the number of experimental samples surpasses the variance of snapshots, which is defined as the shadow norm $||O||_{\text{sh}}^2$ \cite{2020NatPh..16.1050H}.

    \begin{figure}[tb]
        \centering
        \includegraphics[width=0.83\linewidth]{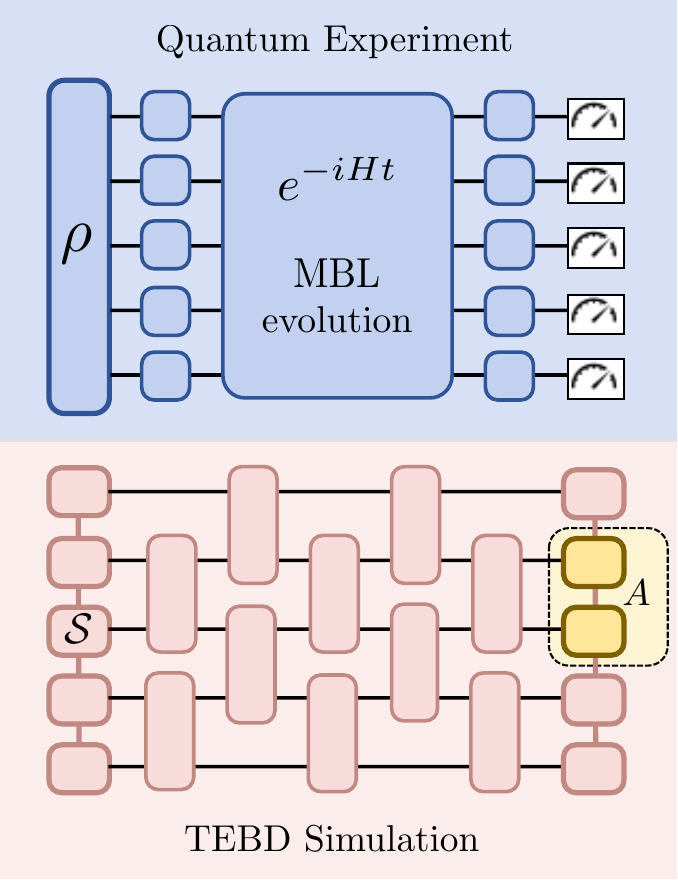}
        \caption{Schematics of our MBL-based classical shadow tomography. In order to predict the properties of a density matrix $\rho$ efficiently, we evolve the system by two layers of random single-qubit gates, separated by an evolution governed by the MBL Hamiltonian. Subsequent measurements are executed on the computational basis. The realization of classical shadow tomography hinges on employing a TEBD simulation, enabling the determination of the inverse channel $\mathcal{M}^{-1}$. }
        \label{fig:schemticas}
    \end{figure}

Different Schemes for classical shadow tomography correspond to different choices of the ensemble $\mathcal{E}_U$. There are two main criteria for useful classical shadow tomography schemes: (i). efficient calculation of $\mathcal{M}^{-1}$ on a classical computer; (ii). low sampling complexity with a small $||O||_{\text{sh}}^2$. In the initial proposal \cite{2020NatPh..16.1050H}, the authors provide examples that include random $N$-qubit Haar random unitaries and tensor products of single-qubit Haar random unitaries\footnote{Here Haar random unitaries can be replaced by Clifford gates, which are 3-designs for qubit systems \cite{2015arXiv151002769W,PhysRevA.96.062336}.} (also known as the Pauli measurement). For these examples, $\mathcal{M}^{-1}$ can be analytically computed. Regarding a Pauli operator with a size of $k$, the shadow norm $||O||_{\text{sh}}^2$ are $2^N$ and $3^k$ respectively. This demonstrates that single-qubit random measurements show better performance for few-body operators. 

Later studies have generalized the protocol to locally scrambled quantum dynamics, first proposed in \cite{PhysRevResearch.5.023027} and then discussed in more general contexts such as Pauli-invariant unitary ensembles \cite{jaffeClassicalShadowsPauliinvariant2024} and the tensor network representation \cite{2023Quant...7.1026A}. The evolution dynamics $U$ has a sandwich structure similar to that in Fig.~\ref{fig:schemticas}, which satisfies $\mathcal{E}_U = \mathcal{E}_{V_iU} = \mathcal{E}_{UV_i}$ for arbitrary single-qubit gate $V_i$. It is also observed that the shadow norm for the locally scrambled ensemble can be related to the operator size distribution $P(n)$ of $UOU^\dagger$, which exhibits different behavior in $N$-qubit Haar random unitaries and tensor products of single-qubit Haar random unitaries \cite{PhysRevLett.130.230403}.
\begin{equation}\label{eq:norm-size}
||O||_{\text{sh}}^2=\Big(\sum_n P(n)3^{-n}\Big)^{-1}.
\end{equation} 
Specifically, in cases where operators are thoroughly scrambled within the $N$-qubit system (as in the scheme with $N$-qubit Haar random unitaries), a typical operator has a size of $O(N)$, resulting in an exponentially large $||O||_{\text{sh}}^2$. On the other hand, local scrambling can reduce $||O||_{\text{sh}}^2$. This reduction occurs due to the decrease in the size of a $k$-local Pauli operator when a $k$-qubit Haar random unitary is applied in the subsystem where the operator $O$ is non-trivial. As a result, in order to attain optimal performance in the measurement of $k$-body Pauli operators, the ensemble $\mathcal{E}_U$ should exhibit information scrambling within small subsystems while remaining unscrambled on larger scales. For instance, dilute shallow circuits, which are constructed using two-qubit gates arranged in a brick-wall architecture decorated by loophole identity gates to slow down the twirling dynamics, are proposed to further reduce the shadow norm to $||O||_{\text{sh}}^2\in (2^k,2.28^k)$ \cite{PhysRevLett.130.230403}. Similar scaling has been observed near the measurement-induced criticality \cite{2023arXiv230715011I,2023arXiv230801653A}.

These schemes rely on the capability to apply random Haar single-qubit gates and realize global Haar randomness via two-qubit gates with high precision. The latter is a task that can be challenging on specific quantum simulators such as neutral atom platforms or trapped ion platforms \cite{zhai2021ultracold}. On the one hand, it is widely agreed that single-qubit Haar unitaries can be and have already been efficiently implemented in experiments \cite{PhysRevA.93.032140,PhysRevResearch.5.023027}. For instance, on the neutral atom platform, single-qubit rotations between two clock states are driven by a two-photon Raman $\Lambda$ transition, utilizing techniques such as composite pulses and dynamical decoupling \cite{kasevichAtomicInterferometryUsing1991,riehleOpticalRamseySpectroscopy1991,levineHighFidelityControlEntanglement2018,levineDispersiveOpticalSystems2022,bluvsteinQuantumProcessorBased2022,bluvsteinLogicalQuantumProcessor2024}.

On the other hand, the realization of two-qubit gates is more challenging. The rapidly evolving development of Rydberg atoms has demonstrated that only certain two-qubit gates, such as CZ gates, can be realized by the Rydberg blockade effect \cite{levineHighFidelityControlEntanglement2018,levineDispersiveOpticalSystems2022,bluvsteinQuantumProcessorBased2022,bluvsteinLogicalQuantumProcessor2024}. Multiple layers of two-qubit evolution require the precise movement of the atoms to the entanglement zone and high fidelity of two-qubit gates. At the current stage, the realization of multiple-layer high-fidelity two-qubit gates is the bottleneck for quantum error correction protocols for NISQ quantum devices. As a result, a recent work proposed a practical algorithm for efficiently realizing classical shadow tomography on the digital gates of the Rydberg atom platform \cite{seifDemonstrationRobustEfficient2024}, which can effectively mitigate additional errors such as qubit crosstalk and measurement error through tensor network-based postprocessing. 

In this study, we propose an alternative scheme that attains high performance for few-body operators, achieved through the many-body localization (MBL) dynamics\footnote{Recently, debates have arisen regarding the existence of MBL in the thermodynamic limit \cite{PhysRevB.95.155129,PhysRevE.102.062144}. In this context, our focus is directed towards its dynamic signature, particularly within systems of moderate size $N$ and evolution time $t$. } \cite{RevModPhys.91.021001,2018CRPhy..19..498A,2015ARCMP...6..383A,2015ARCMP...6...15N,PhysRevLett.111.127201,PhysRevB.90.174202,PhysRevLett.110.067204}, a phenomenon demonstrated in optical lattices \cite{blochObservationManybodyLocalization2015,2016Sci...352.1547C} and trapped ion\cite{2016NatPh..12..907S}. Our scheme is illustrated in Figure \ref{fig:schemticas}. In MBL systems, local integrals of motion (LIOM) exist \cite{PhysRevLett.111.127201,PhysRevB.90.174202,PhysRevLett.110.067204}, effectively preventing systems from undergoing thermalization. However, dephasing still contributes to the scrambling of quantum information, which is probed by the decay of out-of-time-order correlators \cite{Huang:2016knw,Fan:2016ean,2017PhRvB..95f0201S,2017PhRvB..95e4201H,2016arXiv160802765C,2017AnP...52900332C}. We show that our scheme meets the criteria for achieving high efficiency and potentially stands out from other Hamiltonian-driven shadow tomography\cite{PhysRevResearch.4.013054,PhysRevResearch.5.023027,2023Quant...7.1026A,liu2024predictingarbitrarystateproperties}: (i). both $\mathcal{M}$ and its inverse can be computed using the time-evolving block decimation (TEBD) algorithm for large system size $N$ for MBL systems; (ii). the shadow norm $||O||_{\text{sh}}^2$ enters the region $(2^k,2.28^k)$. Therefore, our scheme provides a convincing local Hamiltonian-based approach to storing states obtained by quantum simulations in classical memories.

\section{MBL-based Classical Shadow }\label{section2}
Our work instead provides an efficient realization of the Hamiltonian-driven shadow, which firstly appears at \cite{PhysRevResearch.5.023027}. Previous work requires the analog simulation of Haar random unitaries via certain quantum chaotic Hamiltonians, such as the Rydberg atom Hamiltonian, to satisfy the efficient classical shadow tomography \cite{PhysRevResearch.4.013054,PhysRevResearch.5.023027,2023Quant...7.1026A,liu2024predictingarbitrarystateproperties}. Here, we propose that we could use an MBL-type Hamiltonian to realize the efficient classical shadow protocol with an analog Hamiltonian.

In general, our scheme works for any types of MBL Hamiltonian. The unitary operation $U$ used in the classical shadow protocal is composed of three steps
\begin{equation}\label{eq:unitary_Shadow}
U=(\otimes_i u_i)~e^{-iH t}~(\otimes_j v_j),
\end{equation}
where both $u_i$ and $v_i$ are single-qubit Haar random unitaries, and $H$ belongs to one type of MBL Hamiltonian. From the discussion in Appendix~\ref{app:exp_proposal}, we conclude that $u_i$ and $v_i$ can be robustly implemented and are compatible with MBL dynamics. Therefore, the average over $\mathcal{E}_{U}$ becomes the integration of $u_i$ and $v_i$ under the Haar measure, and we can fix one realization of $H$. For the Hamiltonian evolution part, we choose to adopt the standard random field XXZ model for MBL \cite{RevModPhys.91.021001,2018CRPhy..19..498A}:
\begin{equation}\label{eq:XXZ}
\begin{aligned}
H=&J \sum_i(X_{i}X_{i+1}+Y_{i}Y_{i+1}+\Delta Z_{i}Z_{i+1})\\&+\sum_ih_i Z_i,
\end{aligned}
\end{equation}
where $(X_i,Y_i,Z_i)$ are Pauli operators on site $i$ and $h_i\in [-W,W]$ represents a random magnetic field along the z-direction. In practice, we fix a single disorder realization of $h_i$ when performing the shadow tomography. The system exhibits MBL dynamics for sufficiently large $W/J$ and moderate $\Delta$, where LIOM $\tilde{Z}_i$ emerge, satisfying $[H,\tilde{Z}_i]=0$ \cite{PhysRevLett.111.127201,PhysRevB.90.174202,PhysRevLett.110.067204}. Consequently, the unitary evolution only produces dephasing without any flips of the pseudospins $\tilde{Z}_i$. For $W/J\gg 1$, perturbation theory suggests that we can make the identification $\tilde{Z}_i\approx Z_i$. 

The reasons for introducing singe-qubit Haar random unitaries $v_i$ are twofold. Firstly, the evolution of different Pauli operators ($Z$ versus $X$ and $Y$) exhibits distinct behaviors in the MBL regime due to the significant overlap between $Z$ and LIOM. This discrepancy leads to varying sampling complexities for different $k$-body Pauli operators, which is unfavorable under the "measure first, ask questions later" philosophy \cite{2023NatRP...5....9E}. Moreover, upon the addition of single-qubit Haar random unitaries, $\mathcal{E}_U$ fulfills the conditions for being locally scrambled. This greatly simplifies the calculations involved in determining the inverse channel $\mathcal{M}^{-1}$ and the shadow norm $||O||_{\text{sh}}^2$.

Now we briefly review the classical shadow tomography for locally-scrambled $\mathcal{E}_U$ \cite{PhysRevResearch.5.023027,2023Quant...7.1026A,jaffeClassicalShadowsPauliinvariant2024,PhysRevLett.130.230403}, focusing on predictions of Pauli operators $O$. It is known that Pauli operators are eigenoperators for the quantum channel $\mathcal{M}$. Denoting the corresponding eigenvalue as $\lambda_O$, we have $\langle O\rangle=\lambda_O^{-1}\text{tr}(O\sigma)$. Moreover, the shadow norm $||O||_{\text{sh}}^2$ is also related to $\lambda_O$ by $||O||_{\text{sh}}^2=\lambda_O^{-1}$. The remaining task is to compute $\lambda_O$ 
\begin{equation}\label{eq:lambda}
\begin{aligned}
\lambda_O&=D^{-1}\text{tr}(O\mathcal{M}(O) )
=\mathds{E}_{\mathcal{E}_U}[\langle 0|U OU^\dagger|0\rangle^2]
\end{aligned}
\end{equation}
where $D=2^N$ is the Hilbert space dimension. We have used locally-scrambled property to replace $|b\rangle$ with $|0\rangle$. Using the operator-state mapping $O \rightarrow |O\rangle\rangle$, we have:
\begin{equation}\label{eq:lambda2}
\langle 0|U OU^\dagger|0\rangle^2=D\left(\langle\langle O|U^\dagger\otimes U^T |0\rangle\rangle\right)^2,
\end{equation}
where we have introduced $|0\rangle\rangle\equiv |0\rangle \otimes |0\rangle$. The operator space $|\psi\rangle\rangle$ possesses a dimension of $D^2=4^N$. The detailed operator-state mapping can be found in the appendix \ref{app:op_state}. In express \eqref{eq:lambda2} as a single expectation value, we further extend the notion by introducing a doubled operator space with dimension $D^4=16^N$, e.g. $|0))\equiv |0\rangle\rangle \otimes |0\rangle\rangle$, which gives
\begin{equation}\label{eq:computenorm}
\lambda_O=D~\mathds{E}_{\mathcal{E}_U}[\langle\langle O|^{\otimes 2}(U^\dagger\otimes U^T )^{\otimes 2}|0))].
\end{equation}
To proceed, we carry out the average over $u_i$ and $v_i$ analytically, and the detailed calculation is in appendix \ref{app:haar_average}. The result reads
\begin{equation}
\lambda_O=((A|(e^{iHt}\otimes e^{-iH^Tt} )^{\otimes 2}|\mathcal{S})).
\end{equation}
$|A))$ represents the region basis states, which solely depends on the subregion $A$ on which operator $O$ acts non-trivially. In this work, we adopt the normalization that $|A))=\otimes_i|\psi_i))$, with
\begin{equation}\label{eq:A_op2_state}
|\psi_{i}))= \begin{cases}
      3^{-1}\sum_{P=X,Y,Z}|P\rangle\rangle\otimes|P\rangle\rangle & \text{for $i\in A$,}\\
      |I\rangle\rangle\otimes|I\rangle\rangle & \text{for $i\notin A$.}
    \end{cases}   
\end{equation}
$|\mathcal{S}))=\otimes_i|\mathcal{S}_i)) $ is obtained by averaging over $u_i$:
\begin{equation}\label{eq:S_op2_state}
|\mathcal{S}_i))=|I\rangle\rangle\otimes|I\rangle\rangle+3^{-1}\sum_{P=X,Y,Z}|P\rangle\rangle\otimes|P\rangle\rangle .
\end{equation} 
The state $|\mathcal{S}))$ measures the size of the evolved operator $UOU^\dagger$: any non-trivial Pauli operator contributes a factor of $1/3$. This gives 
\begin{equation}\label{eq:lambda_OpNorm}
    \lambda_O=\sum_n P(n)3^{-n}
\end{equation}
, as stated in the previous section, where the operator size distribution $P(n)$ is averaged over all Pauli operators supported strictly on the subsystem $A$. 

In previous works, where $U$ is composed of locally-scrambled few-body gates, the evolution can be simulated in the subspace of region basis states, which is known as the entanglement feature \cite{2018PhRvB..97d5153Y,2018PhRvB..98a4309Y}. This subspace has a dimension of $D=2^N$. However, this approach is no longer applicable to our Hamiltonian-based scheme. Consequently, we now compute \eqref{eq:computenorm} directly within the doubled operator space, which has a dimension of $D^4=16^N$. Nevertheless, since our Hamiltonian is many-body localized, the simulation can be efficiently executed on a classical computer using a TEBD algorithm \cite{10.21468/SciPostPhysCodeb.4} for $N\sim 10^2$: 
\begin{enumerate}[nosep]
    \item  prepare an initial state $|\mathcal{S}))$ utilizing a matrix product state (MPS) representation with a bond dimension of 1;
\item construct the total Hamiltonian $H_\text{tot}=H_{(1)}-H_{(2)}^T+H_{(3)}-H_{(4)}^T$, where subscripts ${(i)}$ denotes the Hilbert space where $H$ acts non-trivially;
\item Evolve $|\mathcal{S}))$ by $H_\text{tot}$ for a duration of time $t$ through the TEBD algorithm;
\item  Determine the overlap between the resulting $|\mathcal{S}(t)))$ and $|A))$ to derive $\lambda_O$.
\end{enumerate}
This process notably yields $||O||_{\text{sh}}^2$ for any arbitrary Pauli operator $O$ after a single run of the TEBD algorithm. Our algorithm can also be applied to quantum chaotic Hamiltonians for short evolution time. 

\section{Phenomenological Model Analysis}\label{sec:PhenomelogicalModel}
 Deep in the MBL regime, the emergence of LIOM largely constrains the operator dynamics \cite{PhysRevLett.111.127201,PhysRevB.90.174202,PhysRevLett.110.067204}. In this section, we provide an estimation of $||O||_{\text{sh}}^2$ due to the dephasing. Our theoretical tool is the phenomenological model proposed as an effective theory of MBL \cite{PhysRevLett.111.127201,PhysRevB.90.174202,PhysRevLett.110.067204}: 
 \begin{equation}\label{eq:MBLeff}
 H_{\text{eff}}=\sum_i B_i \tilde{Z}_i+\sum_{i<j} J_{ij}\tilde{Z}_i\tilde{Z}_j+...
 \end{equation} 
 $B_i\in [-B,B]$ is a random magnetic field. As this term doesn't create entanglement among distinct pseudospins, it doesn't contribute to the growth of operators. We thus set $B=0$ for conciseness. $J_{ij}=\tilde{J}_{ij}e^{-|i-j|/\xi}$, where $\xi$ is the localization length and $\tilde{J}_{ij}$ are several independent uniform distributions within $[-J_0,J_0]$. In principle, there are higher-order terms in \eqref{eq:MBLeff}, which characterize interactions among more than two pseudospins.
 With the justification provided in Appendix~\ref{supp_ssub:High_order}, we can show that it is reasonable to primarily consider the two-body interaction term, as the higher-order terms do not result in additional contributions when analyzing the shadow norm.

To estimate $\lambda_O$, we evaluate the expansion of a certain Pauli operator $O$ using the effective Hamiltonian given by \eqref{eq:MBLeff}. Our emphasis is on the regime of high $W/J$, within which there's no necessity to differentiate between $Z$ and $\tilde{Z}$. Nonetheless, we anticipate our estimation to remain valid for smaller $W/J$, provided that the size of Pauli operator $k$ significantly exceeds the correlation length $\xi$. 

Before we analyze the general operator $O$, we consider the evolution of $O=\tilde{Z}_1 \tilde{X}_2$, $k=2$ as an illustrative example. A straightforward calculation shows that:
\begin{equation}
\begin{aligned}
e^{-iH_\text{eff}t} &\tilde{Z}_1 \tilde{X}_2e^{iH_\text{eff}t} \\ 
=&\prod_{j\neq 1,2}  (\cos(2J_{2j}t)-i\sin (2J_{2j}t) \tilde{Z}_2\tilde{Z}_j)\\
\times&(\cos(2J_{12}t)\tilde{Z}_1\tilde{X}_2+\sin (2J_{12}t) \tilde{Y}_2),
\end{aligned}
\end{equation}
where the evolution is driven by the interaction between the second spin and other spins, since $\tilde{X}_2$ does not commute with the Hamiltonian. We can then compute the corresponding contribution to $\lambda_O$ simplified in Eq.~\eqref{eq:lambda_OpNorm} as
\begin{equation}\label{eq:reseffnoaverage}
\begin{aligned}
\sum_n P(n)3^{-n}=&\frac{1}{3}\Big(\sin (2J_{12}t)^2+\frac{1}{3}\cos (2J_{12}t)^2 \Big)\\
&\prod_{j\neq 1,2}\Big(\cos (2J_{2j}t)^2+\frac{1}{3}\sin (2J_{2j}t)^2 \Big).
\end{aligned}
\end{equation} 
We first focus on the result after the disorder average over $J_{ij}$, which leads to 
\begin{equation}\label{eq:reseff}
\begin{aligned}
\overline{\sum_n P(n)3^{-n}}=&\frac{1}{3}\Big(\frac{2}{3}-\frac{1}{3}f(4 J_0e^{-1/\xi} t)\Big)\\&\prod_{j\neq 1,2}\Big(\frac{2}{3}+\frac{1}{3}f(4 J_0 e^{-|j-2|/\xi}t)\Big).
\end{aligned}
\end{equation}
Here, the overline means the random averages over an ensemble of $J_{ij}$, and $f(x)=\text{sinc}(x)=\sin(x)/x$ depends on the details of the distribution function of $\tilde{J}_{ij}$. However, the ensuing discussions solely rely on its asymptotic behavior, $f(0)=1$ and $f(\infty)=0$, which are expected to be universal. 

This explicit calculation can be further understood using the universal feature of scrambling physics. We aim to separately discuss the contribution to the $\overline{\sum_n P(n)3^{-n}}$ for different sites. The rationale comes from the assumption that, after scrambling time, $\overline{\sum_n P(n)3^{-n}} = \prod_i \overline{\sum_{n_i} P(n_i)3^{-n_i}}$, where $P(n_i)$ is the operator size distribution function reduced to a single site $i$, i.e., $P(n_i) \equiv \sum_{n \neq n_i} P(n)$. Here, the non-trivial Pauli operator on site $i$ will contribute $1/3$ to the formula, while the trivial Pauli operator will contribute $1$. 
We discuss $\overline{\sum_{n_i} P(n_i)3^{-n_i}}$ for different sites:
\begin{enumerate}[label=(\alph*)]
    \item \emph{The initial operator $O$ supports a non-trivial Pauli operator $\tilde{X}_i$ or $\tilde{Y}_i$ on site $i$. This local single-site operator will equally mix $\tilde{X}_i$ and $\tilde{Y}_i$ after scrambling time.}
    \item \emph{The initial operator $O$ supports a non-trivial Pauli operator $\tilde{Z}_i$ on site $i$. This local single-site operator will equally mix $\tilde{Z}_i$ and $\tilde{I}_i$ after scrambling time.}
    \item \emph{The initial operator $O$ does not support any non-trivial Pauli operator on site $i$. This local single-site operator will equally mix $\tilde{Z}_i$ and $\tilde{I}_i$ after this site enters the light cone of the evolving operator.}
\end{enumerate}
The reason we distinguish cases (a) and (b) is the local $U(1)$ symmetry. Since the Hamiltonian commutes with $\tilde{Z}_i$, the Heisenberg evolution is block diagonalized according to different representations under the local $U(1)$ transformation generated by $\tilde{Z}_i$, which solely mixes $(\tilde{Z}_i, \tilde{I}_i)$ or $(\tilde{X}_i, \tilde{Y}_i)$.

Returning to our example, we can classify all terms in Eq.~\eqref{eq:reseff} into the three cases above. For case (a), site 2 always contains a non-trivial Pauli operator $\tilde{X}_i$ or $\tilde{Y}_i$, yielding a factor of $1/3$ in $\overline{\sum_{n_i} P(n_i)3^{-n_i}}$. Correspondingly, this factor appears in the first line of \eqref{eq:reseff}.

For case (b), the initial operator on site 1 is $\tilde{Z}_i$. However, in the long-time limit, we expect equal probability of finding $\tilde{Z}_1$ or $\tilde{I}_1$, which contributes a factor of $\frac{1}{2} \cdot 3^{-1} + \frac{1}{2} \cdot 1 = \frac{2}{3}$ in $\overline{\sum_{n_i} P(n_i)3^{-n_i}}$. This value is exactly consistent with the first bracket in \eqref{eq:reseff}, after scrambling time so that we can ignore $\frac{1}{3}f(4 J_0 e^{-1/\xi})$.

For case (c), the sites $i \neq 1, 2$ also have the probability of contributing to $\overline{\sum_{n_i} P(n_i)3^{-n_i}}$, since the long-range interaction in Eq.~\eqref{eq:MBLeff} can also lead to equal mixing of $\tilde{Z}_i$ and $\tilde{I}_i$ operators, thus contributing $\frac{1}{2} \cdot 3^{-1} + \frac{1}{2} \cdot 1 = \frac{2}{3}$ to $\overline{\sum_{n_i} P(n_i)3^{-n_i}}$. This corresponds to the second line of \eqref{eq:reseff}, describing the contribution of the remaining sites. However, only the sites in the light cone will contribute to the operator size for a fixed time. It is known that MBL systems exhibit a logarithmic light cone \cite{RevModPhys.91.021001, 2018CRPhy..19..498A}: a signal traverses a distance $\Delta x$ after an evolution time $\Delta t \sim J_0^{-1} e^{\Delta x/\xi}$. Therefore, there are approximate $\Delta N = 2\xi \log (J_0 t)$ sites in the light cone\footnote{Here, the factor of two comes from the two boundaries of the subregion $A$.}. This collectively contributes a factor of $(\frac{2}{3})^{\Delta N} = (\frac{2}{3})^{2 \xi \log(J_0 t)}$.

    \begin{figure}[tb]
        \centering
        \includegraphics[width=0.95\linewidth]{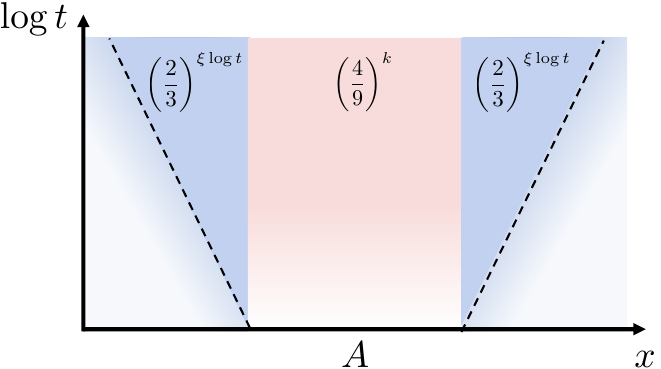}
        \caption{ An illustration for the estimation of $\lambda_O$ in the phenomenological model. The result \eqref{eq:reslambdaeff} contains contributions from the subsystem $A$, which is the support of the Pauli operator, and the remaining qubits in the logarithmic lightcone. }
        \label{fig:phenomenologicalresult}
    \end{figure}

With this understanding, we are now prepared to offer an estimation for $\lambda_O$ by averaging across distinct Pauli operators $O$, all supported strictly within the subsystem $A$ encompassing consecutive $k$ sites. We arrive at the following statement:

\textbf{Statement 1}. \textit{In the long-time limit, where $J_0 e^{-1/\xi} t \gg 1$, the estimation of $\lambda_O$ for any random operator $O$ in subsystem $A$ with length $k$ asymptotically grows according to the formula}
\begin{equation}\label{eq:reslambdaeff}
\lambda_O=(||O||_{\text{sh}}^2)^{-1}=\left(\frac{4}{9}\right)^k\left(\frac{2}{3}\right)^{2\xi \log (J_0 t)}.
\end{equation}

We provide the detailed proof in Appendix~\ref{supp_ssub:Proof_Theorem1}. Here, we illustrate the contribution of $\lambda_O$ in Figure~\ref{fig:phenomenologicalresult}. For a fixed evolution time $t$, our result predicts $||O||_{\text{sh}}^2 \propto 2.25^k$. This falls within the range of optimal scaling achieved by the shallow circuit scheme, which is $||O||_{\text{sh}}^2 \in (2^k, 2.28^k)$ \cite{PhysRevLett.130.230403}. 

We finally comment on the shadow norm $||O||_{\text{sh}}^2$ without the disorder average. Naturally, we anticipate that the outcome will exhibit prolonged oscillations, akin to those in Eq. \eqref{eq:reseffnoaverage}. However, the extended time average at large values of $t J_{ij}$ plays a role equivalent to that of the disorder average in the case of independent $J_{ij}$. Consequently, the non-averaged $||O||_{\text{sh}}^2$ is expected to oscillate around \eqref{eq:reslambdaeff}. Notably, it might exhibit enhanced performance for specific time instances $t$.

\section{Numerical Demonstration}
In this section, we furnish a numerical demonstration showcasing the efficacy of our MBL-based classical shadow tomography, facilitated by the TEBD algorithm. During the sampling and reconstruction process, the single-qubit Haar random gates are sampled by 24 single-qubit Clifford gates. To conduct our simulations, we employ the \texttt{ITensors.jl} package \cite{10.21468/SciPostPhysCodeb.4}. The source code can be found on the website \cite{tgzhou_code}. 

\begin{figure}[tb]
        \centering
        \includegraphics[width=1.0\linewidth]{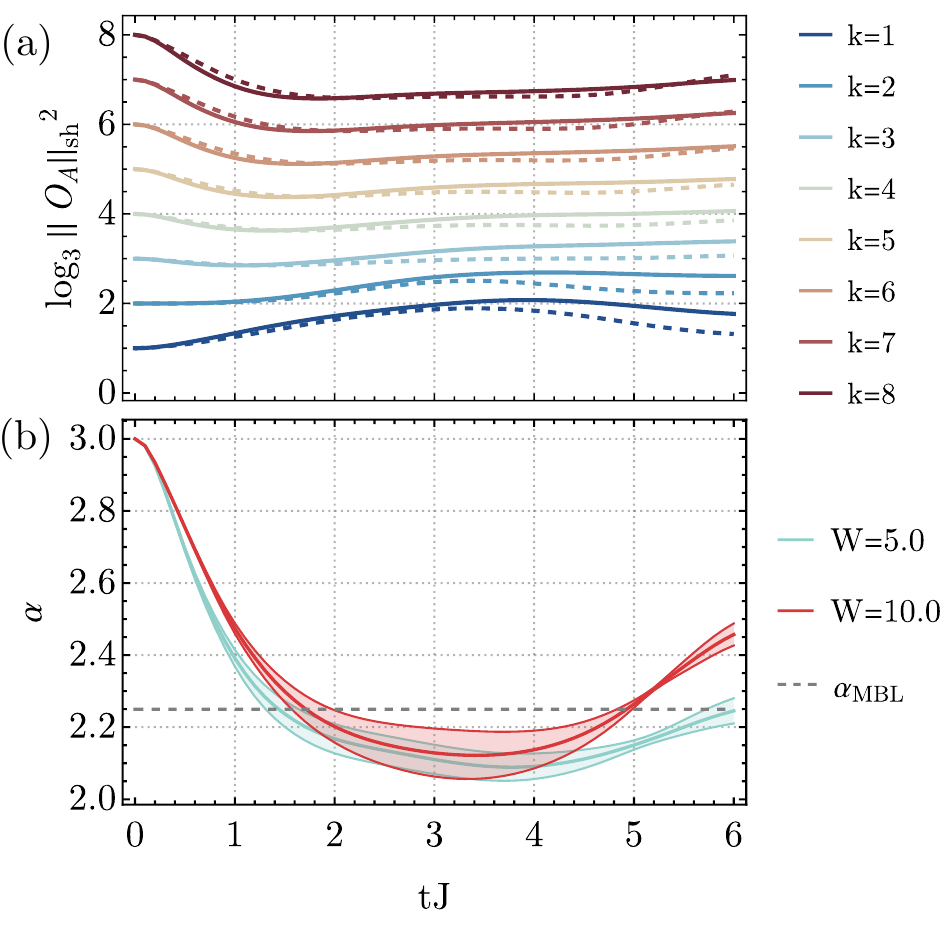}
        \caption{Time-dependent shadow norm obtained from TEBD simulations. We set the system sites to $N=50$ with open boundary conditions and $\Delta=1$. We take each Trotter step $\tau J=0.1$ and a bond dimension of $\chi=100$. We have verified the convergence of our results for different values of $\chi$.
        (a) Shadow norm $||O_A||_{\text{sh}}^2$ averaged over different subsystems. Operators with indices $k=1$ to $8$ are represented by distinct colors. The solid lines correspond to a disorder strength of $W=5.0$, while the dashed lines correspond to $W=10.0$.
        (b) Fitting results of $||O||_{\text{sh}}^2 = c_{0} \alpha^k$. The legend indicates two different disorder strengths, shown by solid lines, with the shaded region representing twice the standard deviation of the fitting parameter. $\alpha_{\text{MBL}}= 9/4$ is the prediction of the phenomenological model after the disorder average.
        }
        \label{fig:OAsh_MBL}
    \end{figure}

In order to predict Pauli observables, we should first compute the shadow norm $||O_A||_{\text{sh}}^2=\lambda_O^{-1}$, as expounded upon in section \ref{section2}. The results for $N=50$ and $\Delta=1$ are presented in Fig.~\ref{fig:OAsh_MBL} (a) after taking $\log_3$. Here, we have computed the average of $||O_A||_{\text{sh}}^2$ across all operators strictly confined within the subsystem $A$, which encompasses consecutive sites within a range of $k$ values ($k=\in\{1,2,...,8\}$). The solid lines correspond to a disorder strength of $W=5.0$, while the dashed lines represent $W=10.0$. We confirm that the numerical result is converged by taking different bond dimensions $\chi=64, 100$ and find no significant change. When $tJ=0$, the results align with those for single-qubit Haar random unitaries, which is $||O_A||_{\text{sh}}^2 = 3^k$. Upon introducing further dynamics through an MBL Hamiltonian evolution, the outcomes bear a resemblance to those of the local-scrambled two-body gates protocol or the Hamiltonian-driven protocol (as referenced in \cite{PhysRevResearch.4.013054,PhysRevResearch.5.023027,2023Quant...7.1026A}), despite the MBL preventing the system from reaching thermal equilibrium.

Nevertheless, it is worth highlighting two key aspects of our protocol. First, the behavior of the shadow norm for small values of $k$ does not consistently involve growth with extended evolution time. This phenomenon finds its explanation in the context of the MBL phenomenon, where the dynamics of operators are confined by the conservation of the LIOMs. This characteristic suggests that the reconstruction of a few body operators becomes more achievable within the quantum simulator framework, as it does not demand precise time control.
Secondly, the dashed line corresponds to $W=10.0$, delving further into the MBL region. Our findings suggest that this results in a reduced shadow norm, particularly pronounced within instances exhibiting stronger MBL. This effect remains significant primarily for operators with small size $k$.

To investigate the $k$ dependence of the shadow norm, we take the following ansatz $||O||_{\text{sh}}^2 = c_{0} \alpha^k$. By taking the logarithm and subsequently performing linear regression, we can deduce the exponent $\alpha$ for varying evolution durations. The results are illustrated in Fig.~\ref{fig:OAsh_MBL} (b), with the shaded region denoting twice the standard deviation. As time increases, $\alpha$ decreases from the single-qubit result where $\alpha=3$, and it enters the optimized scaling region achieved by the shallow circuit scheme, with $\alpha$ ranging within $\left(2, 2.28 \right)$ as reported in \cite{PhysRevLett.130.230403}. It reaches a minimum which is around $\alpha \approx 2.1$, a value significantly smaller than the phenomenological model prediction $\alpha_{\text{MBL}}=9/4$ after performing the disorder average. Subsequently, $\alpha$ increases and exhibits oscillations around $\alpha_{\text{MBL}}$, aligning with the discussions presented in the preceding section.

\begin{figure}[tb]
        \centering
        \includegraphics[width=1.0\linewidth]{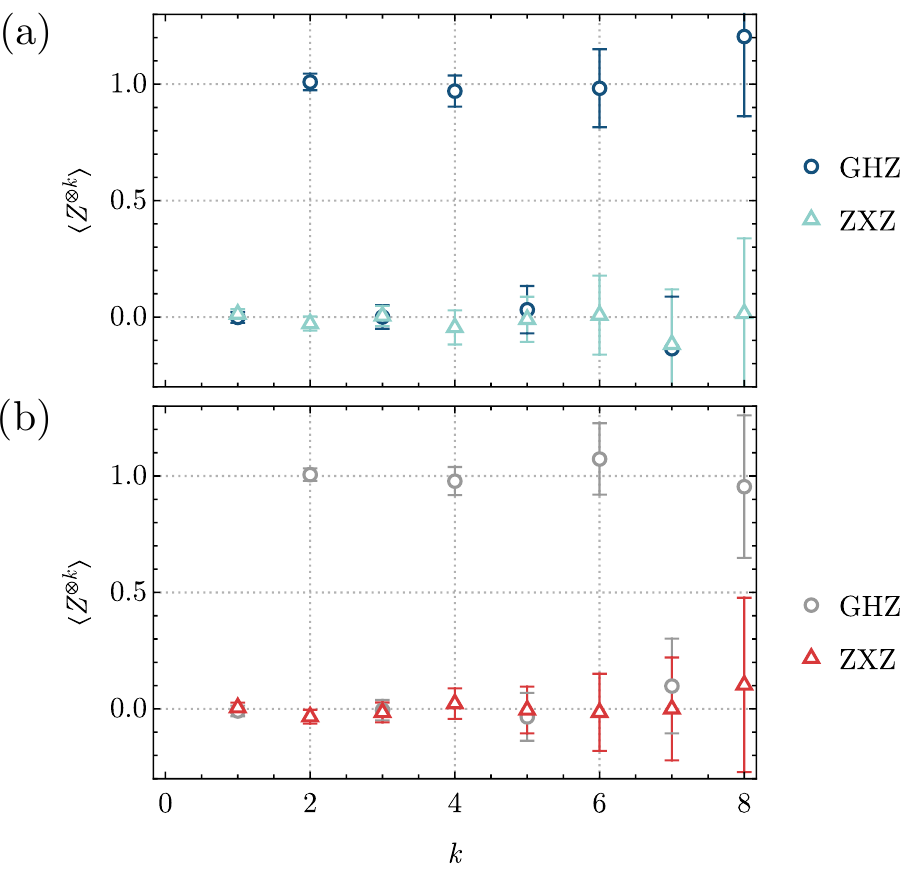}
        \caption{The estimated Pauli observable $Z^{\otimes k} \equiv \bigotimes_{i=N/2+1}^{N/2 + k} Z_i$ with different operator length $k$. The initial state $\rho$ corresponds to both open boundary conditions GHZ states (circles) and ZXZ states (triangles). These predictions are generated from 50,000 measurement samples obtained through TEBD simulations with system size $N=30$, $\Delta=1$, trotter steps $\tau J=0.1$, and a total evolution time of $tJ=2.0$. The error bars represent two standard deviations from the estimated values. Panels (a) and (b) correspond to disorder strengths of $W=5.0$ and $W=10.0$, respectively. }
        \label{fig:OAexpect_MBL}
\end{figure}

We further validate the feasibility and efficiency of our MBL-based scheme by conducting direct simulations to reconstruct Pauli observables using measurement outcomes generated by TEBD simulations. We contemplate two illustrative initial states: the Greenberger-Horne-Zeilinger (GHZ) state denoted as $\rho_{\text{GHZ}}$ and the ZXZ state denoted as $\rho_{\text{ZXZ}}$. Both states can be characterized by their respective stabilizer groups:
\begin{equation}
    \begin{split}
        \mathcal{G}_{\text{GHZ}} &= \langle Z_1 Z_2, \cdots Z_{n-1}Z_n, \Pi_{i=1}^n X_i \rangle, \\
        \mathcal{G}_{\text{ZXZ}} &= \langle Z_1 X_2 Z_3, \cdots , Z_n X_1 Z_2\rangle. \\
    \end{split}
\end{equation}
In the process of numerical construction, we employ the precise MPS representation of the GHZ state under open boundary conditions (OBC). As for the ZXZ state, we iteratively apply the stabilizer projection operator $\prod_i (\mathds{I} + Z_i X_{i+1} Z_{i+2})/{2}$ to an arbitrary OBC state which is not orthogonal to the ZXZ state, such as the state where all spins are oriented upwards.
We measure the Pauli observable $Z^{\otimes k} \equiv \bigotimes_{i=N/2- \lfloor (k-1)/2 \rfloor}^{N/2 + \lfloor k/2 \rfloor} Z_i$  for both the GHZ and ZXZ states. In this context, we select the Pauli string located at the system's center to mitigate potential boundary effects arising from the use of OBC. The theoretical results read
\begin{equation}
    \begin{split}
        \langle \text{GHZ} | Z^{\otimes k} | \text{GHZ} \rangle &= \left( (-1)^k + 1 \right)/2, \\
        \langle \text{ZXZ} | Z^{\otimes k} | \text{ZXZ} \rangle &= 0. \\
    \end{split}
\end{equation}

\begin{figure}[tb]
        \centering
        \includegraphics[width=0.95\linewidth]{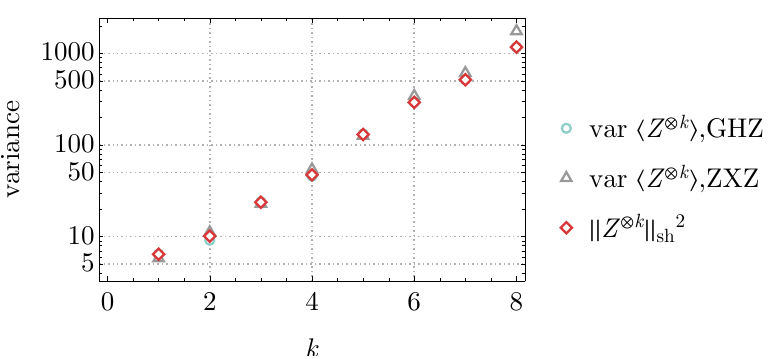}
        \caption{A comparison between the variances over all samples and the shadow norm $||O_A||_{\text{sh}}^2$. The variances in the GHZ or the ZXZ state are extracted from the same data which produces Fig.~\ref{fig:OAexpect_MBL}(a). The results demonstrate the shadow norm provides an accurate prediction for the variance of the classical snapshots. }
        \label{fig:OAvar_MBL}
\end{figure}

According to Fig.~\ref{fig:OAsh_MBL}(b), the shadow norm for $tJ=2.0$ already enters the range of optimal scaling achieved by the shallow circuit scheme. Consequently, we maintain the Hamiltonian evolution for a fixed total time of $tJ=2.0$. To ensure an ample sample size, we opt for a relatively modest system size of $N=30$. During the shadow norm simulation step, we set the MPS bond dimension as $\chi=200$. Correspondingly, in generating samples through TEBD simulations, we fix the MPS cutoff as $\delta_{\chi}=10^{-14}$. The mean value of the simulated classical snapshots leads to the non-biased expectation value of Pauli observables but with error bars. By definition, the variance of the classical snapshots $\sigma_m$ is defined by $$\text{var } \langle O \rangle \equiv \frac{1}{M} \sum_{m=1}^M \left[\Tr\left( \mathcal{M}^{-1} (O) \sigma_m\right) - \langle O \rangle \right]^2$$. The standard deviation to the expectation value is $\sqrt{\text{var } \langle O \rangle / M}$, where $M=50000$ is the measurement samples. As shown in Fig.~\ref{fig:OAexpect_MBL}, the prediction of Pauli observables agrees well with the theoretical analysis within the error bars.

Since the variance directly reveals the efficiency of the protocol, we compare the variance with the shadow norm directly obtained through the TEBD simulation. The numerical results in Fig.~\ref{fig:OAvar_MBL} indicate that the variance is not dependent on the initial states. This consistency aligns with our framework for obtaining the shadow norm. Furthermore, the variance from both initial states aligns well with the simulated shadow norm, which demonstrates the accuracy of our numerical simulations. Moreover, the linearity of the shadow norm in the logarithmic plot provides further support for the reasonability of the exponential ansatz for the shadow norm. 

\section{Discussion}
In this work, we introduce an MBL-based scheme designed for conducting classical shadow tomography in two promising quantum platforms, including neutral atoms and trapped ions. The system undergoes evolution through two layers of single-qubit random unitaries, interspersed with evolution governed by the MBL Hamiltonian. Utilizing a phenomenological model, our estimation indicates that the dephasing of LIOMs leads to a disorder-averaged shadow norm $||O||_{\text{sh}}^2$ scaling as $(9/4)^k$, accompanied by a correction induced by the logarithmic lightcone. Without the disorder average, we anticipate persistent oscillations of $||O||_{\text{sh}}^2$ around $(9/4)^k$, showcasing enhanced performance during specific time instances. These theoretical predictions find validation through TEBD simulations, which demonstrate the efficiency of our proposed scheme.

There are several interesting directions. Firstly, recent research \cite{2023arXiv230715011I, 2023arXiv230801653A} underscores that measurement-induced criticality achieves tomographic optimality. Given the analogy between measurement-induced phase transitions and the transition from chaos to non-chaos in MBL systems, a natural inquiry arises: does the optimal $\alpha$ materialize around the localization transition? Interestingly, our findings in Figure \ref{fig:OAsh_MBL} suggest that the optical $\alpha$ for $W=5$ is smaller than for $W=10$. However, a meticulous exploration of this matter necessitates the development of an efficient algorithm for computing the shadow norm $||O||_{\text{sh}}^2$ within thermalized systems. Secondly, the prevention of thermalization in MBL systems stems from the emergence of LIOMs. Yet, there exist other system types that exhibit non-thermal quantum dynamics, such as 1D integrable systems and systems with fragmented Hilbert spaces. Investigating their effectiveness in classical shadow tomography presents another captivating avenue of exploration. The answers to these questions are deferred to future research endeavors.

\emph{Acknowledgement.} 
We thank Xiao-Liang Qi, Yi-Zhuang You, and Zhao-Yi Zeng for their helpful discussions. The project is supported by NSFC under Grant No. 12374477.

\bibliographystyle{unsrtnat}
\bibliography{ref.bib}

\begin{thebibliography}{79}
\providecommand{\natexlab}[1]{#1}
\providecommand{\url}[1]{\texttt{#1}}
\expandafter\ifx\csname urlstyle\endcsname\relax
  \providecommand{\doi}[1]{doi: #1}\else
  \providecommand{\doi}{doi: \begingroup \urlstyle{rm}\Url}\fi

\bibitem[Arute et~al.(2019)]{Arute:2019zxq}
Frank Arute et~al.
\newblock {Quantum supremacy using a programmable superconducting processor}.
\newblock \emph{Nature}, 574\penalty0 (7779):\penalty0 505--510, 2019.
\newblock \doi{10.1038/s41586-019-1666-5}.

\bibitem[Altman et~al.(2021)]{PRXQuantum.2.017003}
Ehud Altman et~al.
\newblock Quantum simulators: Architectures and opportunities.
\newblock \emph{PRX Quantum}, 2:\penalty0 017003, Feb 2021.
\newblock \doi{10.1103/PRXQuantum.2.017003}.

\bibitem[Zhong et~al.(2021)]{PhysRevLett.127.180502}
Han-Sen Zhong et~al.
\newblock Phase-programmable gaussian boson sampling using stimulated squeezed
  light.
\newblock \emph{Phys. Rev. Lett.}, 127:\penalty0 180502, Oct 2021.
\newblock \doi{10.1103/PhysRevLett.127.180502}.

\bibitem[Wu et~al.(2021)]{PhysRevLett.127.180501}
Yulin Wu et~al.
\newblock Strong quantum computational advantage using a superconducting
  quantum processor.
\newblock \emph{Phys. Rev. Lett.}, 127:\penalty0 180501, Oct 2021.
\newblock \doi{10.1103/PhysRevLett.127.180501}.

\bibitem[Mi et~al.(2021)]{Mi:2021gdf}
Xiao Mi et~al.
\newblock {Information scrambling in quantum circuits}.
\newblock \emph{Science}, 374\penalty0 (6574):\penalty0 abg5029, 2021.
\newblock \doi{10.1126/science.abg5029}.

\bibitem[Ebadi et~al.(2021)]{Ebadi:2020ldi}
Sepehr Ebadi et~al.
\newblock {Quantum phases of matter on a 256-atom programmable quantum
  simulator}.
\newblock \emph{Nature}, 595\penalty0 (7866):\penalty0 227--232, 2021.
\newblock \doi{10.1038/s41586-021-03582-4}.

\bibitem[Semeghini et~al.(2021)]{Semeghini:2021wls}
Giulia Semeghini et~al.
\newblock {Probing topological spin liquids on a programmable quantum
  simulator}.
\newblock \emph{Science}, 374\penalty0 (6572):\penalty0 abi8794, 2021.
\newblock \doi{10.1126/science.abi8794}.

\bibitem[{Zhu} et~al.(2022)]{2022SciBu..67..240Z}
Qingling {Zhu} et~al.
\newblock {Quantum computational advantage via 60-qubit 24-cycle random circuit
  sampling}.
\newblock \emph{Science Bulletin}, 67\penalty0 (3):\penalty0 240--245, February
  2022.
\newblock \doi{10.1016/j.scib.2021.10.017}.

\bibitem[{Flammia} et~al.(2012){Flammia}, {Gross}, {Liu}, and
  {Eisert}]{2012NJPh...14i5022F}
Steven~T. {Flammia}, David {Gross}, Yi-Kai {Liu}, and Jens {Eisert}.
\newblock {Quantum tomography via compressed sensing: error bounds, sample
  complexity and efficient estimators}.
\newblock \emph{New Journal of Physics}, 14\penalty0 (9):\penalty0 095022,
  September 2012.
\newblock \doi{10.1088/1367-2630/14/9/095022}.

\bibitem[{O'Donnell} and {Wright}(2015)]{2015arXiv150801907O}
Ryan {O'Donnell} and John {Wright}.
\newblock {Efficient quantum tomography}.
\newblock \emph{arXiv e-prints}, art. arXiv:1508.01907, August 2015.
\newblock \doi{10.48550/arXiv.1508.01907}.

\bibitem[{Haah} et~al.(2015){Haah}, {Harrow}, {Ji}, {Wu}, and
  {Yu}]{2015arXiv150801797H}
Jeongwan {Haah}, Aram~W. {Harrow}, Zhengfeng {Ji}, Xiaodi {Wu}, and Nengkun
  {Yu}.
\newblock {Sample-optimal tomography of quantum states}.
\newblock \emph{arXiv e-prints}, art. arXiv:1508.01797, August 2015.
\newblock \doi{10.48550/arXiv.1508.01797}.

\bibitem[{Aaronson}(2017)]{2017arXiv171101053A}
Scott {Aaronson}.
\newblock {Shadow Tomography of Quantum States}.
\newblock \emph{arXiv e-prints}, art. arXiv:1711.01053, November 2017.
\newblock \doi{10.48550/arXiv.1711.01053}.

\bibitem[{Huang} et~al.(2020){Huang}, {Kueng}, and
  {Preskill}]{2020NatPh..16.1050H}
Hsin-Yuan {Huang}, Richard {Kueng}, and John {Preskill}.
\newblock {Predicting many properties of a quantum system from very few
  measurements}.
\newblock \emph{Nature Physics}, 16\penalty0 (10):\penalty0 1050--1057, June
  2020.
\newblock \doi{10.1038/s41567-020-0932-7}.

\bibitem[{Paini} and {Kalev}(2019)]{2019arXiv191010543P}
Marco {Paini} and Amir {Kalev}.
\newblock {An approximate description of quantum states}.
\newblock \emph{arXiv e-prints}, art. arXiv:1910.10543, October 2019.
\newblock \doi{10.48550/arXiv.1910.10543}.

\bibitem[Chen et~al.(2021)Chen, Yu, Zeng, and Flammia]{PRXQuantum.2.030348}
Senrui Chen, Wenjun Yu, Pei Zeng, and Steven~T. Flammia.
\newblock Robust shadow estimation.
\newblock \emph{PRX Quantum}, 2:\penalty0 030348, Sep 2021.
\newblock \doi{10.1103/PRXQuantum.2.030348}.

\bibitem[{Acharya} et~al.(2021){Acharya}, {Saha}, and
  {Sengupta}]{2021arXiv210505992A}
Atithi {Acharya}, Siddhartha {Saha}, and Anirvan~M. {Sengupta}.
\newblock {Informationally complete POVM-based shadow tomography}.
\newblock \emph{arXiv e-prints}, art. arXiv:2105.05992, May 2021.
\newblock \doi{10.48550/arXiv.2105.05992}.

\bibitem[Struchalin et~al.(2021)Struchalin, Zagorovskii, Kovlakov, Straupe, and
  Kulik]{PRXQuantum.2.010307}
G.I. Struchalin, Ya.~A. Zagorovskii, E.V. Kovlakov, S.S. Straupe, and S.P.
  Kulik.
\newblock Experimental estimation of quantum state properties from classical
  shadows.
\newblock \emph{PRX Quantum}, 2:\penalty0 010307, Jan 2021.
\newblock \doi{10.1103/PRXQuantum.2.010307}.

\bibitem[Levy et~al.(2024)Levy, Luo, and Clark]{2021arXiv211002965L}
Ryan Levy, Di~Luo, and Bryan~K. Clark.
\newblock Classical shadows for quantum process tomography on near-term quantum
  computers.
\newblock \emph{Physical Review Research}, 6\penalty0 (1), January 2024.
\newblock ISSN 2643-1564.
\newblock \doi{10.1103/physrevresearch.6.013029}.

\bibitem[Zhao et~al.(2021)Zhao, Rubin, and Miyake]{PhysRevLett.127.110504}
Andrew Zhao, Nicholas~C. Rubin, and Akimasa Miyake.
\newblock Fermionic partial tomography via classical shadows.
\newblock \emph{Phys. Rev. Lett.}, 127:\penalty0 110504, Sep 2021.
\newblock \doi{10.1103/PhysRevLett.127.110504}.

\bibitem[Wan et~al.(2023)Wan, Huggins, Lee, and
  Babbush]{BabbushMatchgateShadowsFermionic2023}
Kianna Wan, William~J. Huggins, Joonho Lee, and Ryan Babbush.
\newblock Matchgate {{Shadows}} for {{Fermionic Quantum Simulation}}.
\newblock \emph{Commun. Math. Phys.}, 404\penalty0 (2):\penalty0 629--700,
  2023.
\newblock ISSN 1432-0916.
\newblock \doi{10.1007/s00220-023-04844-0}.

\bibitem[Huang et~al.(2022)Huang, Kueng, Torlai, Albert, and
  Preskill]{2021arXiv210612627H}
Hsin-Yuan Huang, Richard Kueng, Giacomo Torlai, Victor~V. Albert, and John
  Preskill.
\newblock Provably efficient machine learning for quantum many-body problems.
\newblock \emph{Science}, 377\penalty0 (6613), September 2022.
\newblock ISSN 1095-9203.
\newblock \doi{10.1126/science.abk3333}.

\bibitem[Hu and You(2022)]{PhysRevResearch.4.013054}
Hong-Ye Hu and Yi-Zhuang You.
\newblock Hamiltonian-driven shadow tomography of quantum states.
\newblock \emph{Phys. Rev. Res.}, 4:\penalty0 013054, Jan 2022.
\newblock \doi{10.1103/PhysRevResearch.4.013054}.

\bibitem[Hu et~al.(2023)Hu, Choi, and You]{PhysRevResearch.5.023027}
Hong-Ye Hu, Soonwon Choi, and Yi-Zhuang You.
\newblock Classical shadow tomography with locally scrambled quantum dynamics.
\newblock \emph{Phys. Rev. Res.}, 5:\penalty0 023027, Apr 2023.
\newblock \doi{10.1103/PhysRevResearch.5.023027}.

\bibitem[{Akhtar} et~al.(2023{\natexlab{a}}){Akhtar}, {Hu}, and
  {You}]{2023Quant...7.1026A}
Ahmed~A. {Akhtar}, Hong-Ye {Hu}, and Yi-Zhuang {You}.
\newblock {Scalable and Flexible Classical Shadow Tomography with Tensor
  Networks}.
\newblock \emph{Quantum}, 7:\penalty0 1026, June 2023{\natexlab{a}}.
\newblock \doi{10.22331/q-2023-06-01-1026}.

\bibitem[Bu et~al.(2024)Bu, Koh, Garcia, and
  Jaffe]{jaffeClassicalShadowsPauliinvariant2024}
Kaifeng Bu, Dax~Enshan Koh, Roy~J. Garcia, and Arthur Jaffe.
\newblock Classical shadows with {{Pauli-invariant}} unitary ensembles.
\newblock \emph{npj Quantum Inf}, 10\penalty0 (1):\penalty0 1--7, January 2024.
\newblock ISSN 2056-6387.
\newblock \doi{10.1038/s41534-023-00801-w}.

\bibitem[Ippoliti et~al.(2023)Ippoliti, Li, Rakovszky, and
  Khemani]{PhysRevLett.130.230403}
Matteo Ippoliti, Yaodong Li, Tibor Rakovszky, and Vedika Khemani.
\newblock Operator relaxation and the optimal depth of classical shadows.
\newblock \emph{Phys. Rev. Lett.}, 130:\penalty0 230403, Jun 2023.
\newblock \doi{10.1103/PhysRevLett.130.230403}.

\bibitem[Bertoni et~al.(2024)Bertoni, Haferkamp, Hinsche, Ioannou, Eisert, and
  Pashayan]{2022arXiv220912924B}
Christian Bertoni, Jonas Haferkamp, Marcel Hinsche, Marios Ioannou, Jens
  Eisert, and Hakop Pashayan.
\newblock Shallow shadows: Expectation estimation using low-depth random
  clifford circuits.
\newblock \emph{Physical Review Letters}, 133\penalty0 (2), July 2024.
\newblock ISSN 1079-7114.
\newblock \doi{10.1103/physrevlett.133.020602}.

\bibitem[Arienzo et~al.(2023)Arienzo, Heinrich, Roth, and
  Kliesch]{2022arXiv221109835A}
Mirko Arienzo, Markus Heinrich, Ingo Roth, and Martin Kliesch.
\newblock Closed-form analytic expressions for shadow estimation with brickwork
  circuits.
\newblock \emph{Quantum Information and Computation}, 23\penalty0 (11,
  12):\penalty0 961–993, September 2023.
\newblock ISSN 1533-7146.
\newblock \doi{10.26421/qic23.11-12-5}.

\bibitem[Ippoliti and Khemani(2024)]{2023arXiv230715011I}
Matteo Ippoliti and Vedika Khemani.
\newblock Learnability transitions in monitored quantum dynamics via
  eavesdropper’s classical shadows.
\newblock \emph{PRX Quantum}, 5\penalty0 (2), April 2024.
\newblock ISSN 2691-3399.
\newblock \doi{10.1103/prxquantum.5.020304}.

\bibitem[{Akhtar} et~al.(2023{\natexlab{b}}){Akhtar}, {Hu}, and
  {You}]{2023arXiv230801653A}
Ahmed~A. {Akhtar}, Hong-Ye {Hu}, and Yi-Zhuang {You}.
\newblock {Measurement-Induced Criticality is Tomographically Optimal}.
\newblock \emph{arXiv e-prints}, art. arXiv:2308.01653, August
  2023{\natexlab{b}}.
\newblock \doi{10.48550/arXiv.2308.01653}.

\bibitem[Koh and Grewal(2022)]{Koh2022classicalshadows}
Dax~Enshan Koh and Sabee Grewal.
\newblock Classical {S}hadows {W}ith {N}oise.
\newblock \emph{{Quantum}}, 6:\penalty0 776, August 2022.
\newblock ISSN 2521-327X.
\newblock \doi{10.22331/q-2022-08-16-776}.

\bibitem[Zhan et~al.(2024)Zhan, Elben, Huang, and
  Tong]{TongLearningConservationLaws2023}
Yongtao Zhan, Andreas Elben, Hsin-Yuan Huang, and Yu~Tong.
\newblock Learning conservation laws in unknown quantum dynamics.
\newblock \emph{PRX Quantum}, 5:\penalty0 010350, Mar 2024.
\newblock \doi{10.1103/PRXQuantum.5.010350}.

\bibitem[Qi et~al.(2019)Qi, Davis, Periwal, and Schleier-Smith]{Qi:2019rpi}
Xiao-Liang Qi, Emily~J. Davis, Avikar Periwal, and Monika Schleier-Smith.
\newblock {Measuring operator size growth in quantum quench experiments}.
\newblock \emph{arXiv e-prints}, art. arXiv:1906.00524, 6 2019.
\newblock \doi{10.48550/arXiv.1906.00524}.

\bibitem[Elben et~al.(2019)Elben, Vermersch, Roos, and
  Zoller]{PhysRevA.99.052323}
A.~Elben, B.~Vermersch, C.~F. Roos, and P.~Zoller.
\newblock Statistical correlations between locally randomized measurements: A
  toolbox for probing entanglement in many-body quantum states.
\newblock \emph{Phys. Rev. A}, 99:\penalty0 052323, May 2019.
\newblock \doi{10.1103/PhysRevA.99.052323}.

\bibitem[{Brydges} et~al.(2019){Brydges}, {Elben}, {Jurcevic}, {Vermersch},
  {Maier}, {Lanyon}, {Zoller}, {Blatt}, and {Roos}]{2019Sci...364..260B}
Tiff {Brydges}, Andreas {Elben}, Petar {Jurcevic}, Beno{\^\i}t {Vermersch},
  Christine {Maier}, Ben~P. {Lanyon}, Peter {Zoller}, Rainer {Blatt}, and
  Christian~F. {Roos}.
\newblock {Probing R{\'e}nyi entanglement entropy via randomized measurements}.
\newblock \emph{Science}, 364\penalty0 (6437):\penalty0 260--263, April 2019.
\newblock \doi{10.1126/science.aau4963}.

\bibitem[{Elben} et~al.(2023){Elben}, {Flammia}, {Huang}, {Kueng}, {Preskill},
  {Vermersch}, and {Zoller}]{2023NatRP...5....9E}
Andreas {Elben}, Steven~T. {Flammia}, Hsin-Yuan {Huang}, Richard {Kueng}, John
  {Preskill}, Beno{\^\i}t {Vermersch}, and Peter {Zoller}.
\newblock {The randomized measurement toolbox}.
\newblock \emph{Nature Reviews Physics}, 5\penalty0 (1):\penalty0 9--24,
  January 2023.
\newblock \doi{10.1038/s42254-022-00535-2}.

\bibitem[Webb(2016)]{2015arXiv151002769W}
Zak Webb.
\newblock The clifford group forms a unitary 3-design.
\newblock \emph{Quantum Info. Comput.}, 16\penalty0 (15–16):\penalty0
  1379–1400, nov 2016.
\newblock ISSN 1533-7146.
\newblock \doi{10.48550/arXiv.1510.02769}.

\bibitem[Zhu(2017)]{PhysRevA.96.062336}
Huangjun Zhu.
\newblock Multiqubit clifford groups are unitary 3-designs.
\newblock \emph{Phys. Rev. A}, 96:\penalty0 062336, Dec 2017.
\newblock \doi{10.1103/PhysRevA.96.062336}.

\bibitem[Zhai(2021)]{zhai2021ultracold}
Hui Zhai.
\newblock \emph{Ultracold Atomic Physics}.
\newblock Cambridge University Press, 2021.
\newblock \doi{10.1017/9781108595216}.

\bibitem[Ma et~al.(2016)Ma, Jackson, Zhou, Chen, Lu, Mazurek, Fisher, Peng,
  Kribs, Resch, Ji, Zeng, and Laflamme]{PhysRevA.93.032140}
Xian Ma, Tyler Jackson, Hui Zhou, Jianxin Chen, Dawei Lu, Michael~D. Mazurek,
  Kent A.~G. Fisher, Xinhua Peng, David Kribs, Kevin~J. Resch, Zhengfeng Ji,
  Bei Zeng, and Raymond Laflamme.
\newblock Pure-state tomography with the expectation value of pauli operators.
\newblock \emph{Phys. Rev. A}, 93:\penalty0 032140, Mar 2016.
\newblock \doi{10.1103/PhysRevA.93.032140}.

\bibitem[Kasevich and Chu(1991)]{kasevichAtomicInterferometryUsing1991}
Mark Kasevich and Steven Chu.
\newblock Atomic interferometry using stimulated {{Raman}} transitions.
\newblock \emph{Phys. Rev. Lett.}, 67\penalty0 (2):\penalty0 181--184, July
  1991.
\newblock \doi{10.1103/PhysRevLett.67.181}.

\bibitem[Riehle et~al.(1991)Riehle, Kisters, Witte, Helmcke, and
  Bord{\'e}]{riehleOpticalRamseySpectroscopy1991}
F.~Riehle, {\relax Th}.~Kisters, A.~Witte, J.~Helmcke, and {\relax Ch}.~J.
  Bord{\'e}.
\newblock Optical {{Ramsey}} spectroscopy in a rotating frame: {{Sagnac}}
  effect in a matter-wave interferometer.
\newblock \emph{Phys. Rev. Lett.}, 67\penalty0 (2):\penalty0 177--180, July
  1991.
\newblock \doi{10.1103/PhysRevLett.67.177}.

\bibitem[Levine et~al.(2018)Levine, Keesling, Omran, Bernien, Schwartz, Zibrov,
  Endres, Greiner, Vuleti{\'c}, and
  Lukin]{levineHighFidelityControlEntanglement2018}
Harry Levine, Alexander Keesling, Ahmed Omran, Hannes Bernien, Sylvain
  Schwartz, Alexander~S. Zibrov, Manuel Endres, Markus Greiner, Vladan
  Vuleti{\'c}, and Mikhail~D. Lukin.
\newblock High-{{Fidelity Control}} and {{Entanglement}} of {{Rydberg-Atom
  Qubits}}.
\newblock \emph{Phys. Rev. Lett.}, 121\penalty0 (12):\penalty0 123603,
  September 2018.
\newblock \doi{10.1103/PhysRevLett.121.123603}.

\bibitem[Levine et~al.(2022)Levine, Bluvstein, Keesling, Wang, Ebadi,
  Semeghini, Omran, Greiner, Vuleti{\'c}, and
  Lukin]{levineDispersiveOpticalSystems2022}
Harry Levine, Dolev Bluvstein, Alexander Keesling, Tout~T. Wang, Sepehr Ebadi,
  Giulia Semeghini, Ahmed Omran, Markus Greiner, Vladan Vuleti{\'c}, and
  Mikhail~D. Lukin.
\newblock Dispersive optical systems for scalable {{Raman}} driving of
  hyperfine qubits.
\newblock \emph{Phys. Rev. A}, 105\penalty0 (3):\penalty0 032618, March 2022.
\newblock \doi{10.1103/PhysRevA.105.032618}.

\bibitem[Bluvstein et~al.(2022)Bluvstein, Levine, Semeghini, Wang, Ebadi,
  Kalinowski, Keesling, Maskara, Pichler, Greiner, Vuleti{\'c}, and
  Lukin]{bluvsteinQuantumProcessorBased2022}
Dolev Bluvstein, Harry Levine, Giulia Semeghini, Tout~T. Wang, Sepehr Ebadi,
  Marcin Kalinowski, Alexander Keesling, Nishad Maskara, Hannes Pichler, Markus
  Greiner, Vladan Vuleti{\'c}, and Mikhail~D. Lukin.
\newblock A quantum processor based on coherent transport of entangled atom
  arrays.
\newblock \emph{Nature}, 604\penalty0 (7906):\penalty0 451--456, April 2022.
\newblock ISSN 1476-4687.
\newblock \doi{10.1038/s41586-022-04592-6}.

\bibitem[Bluvstein et~al.(2024)Bluvstein, Evered, Geim, Li, Zhou, Manovitz,
  Ebadi, Cain, Kalinowski, Hangleiter, Bonilla~Ataides, Maskara, Cong, Gao,
  Sales~Rodriguez, Karolyshyn, Semeghini, Gullans, Greiner, Vuleti{\'c}, and
  Lukin]{bluvsteinLogicalQuantumProcessor2024}
Dolev Bluvstein, Simon~J. Evered, Alexandra~A. Geim, Sophie~H. Li, Hengyun
  Zhou, Tom Manovitz, Sepehr Ebadi, Madelyn Cain, Marcin Kalinowski, Dominik
  Hangleiter, J.~Pablo Bonilla~Ataides, Nishad Maskara, Iris Cong, Xun Gao,
  Pedro Sales~Rodriguez, Thomas Karolyshyn, Giulia Semeghini, Michael~J.
  Gullans, Markus Greiner, Vladan Vuleti{\'c}, and Mikhail~D. Lukin.
\newblock Logical quantum processor based on reconfigurable atom arrays.
\newblock \emph{Nature}, 626\penalty0 (7997):\penalty0 58--65, February 2024.
\newblock ISSN 1476-4687.
\newblock \doi{10.1038/s41586-023-06927-3}.

\bibitem[Hu et~al.(2024)Hu, Gu, Majumder, Ren, Zhang, Wang, You, Minev, Yelin,
  and Seif]{seifDemonstrationRobustEfficient2024}
Hong-Ye Hu, Andi Gu, Swarnadeep Majumder, Hang Ren, Yipei Zhang, Derek~S. Wang,
  Yi-Zhuang You, Zlatko Minev, Susanne~F. Yelin, and Alireza Seif.
\newblock Demonstration of {{Robust}} and {{Efficient Quantum Property
  Learning}} with {{Shallow Shadows}}.
\newblock \penalty0 (arXiv:2402.17911), February 2024.
\newblock \doi{10.48550/arXiv.2402.17911}.

\bibitem[De~Roeck and Huveneers(2017)]{PhysRevB.95.155129}
Wojciech De~Roeck and Fran\ifmmode \mbox{\c{c}}\else~\c{c}\fi{}ois Huveneers.
\newblock Stability and instability towards delocalization in many-body
  localization systems.
\newblock \emph{Phys. Rev. B}, 95:\penalty0 155129, Apr 2017.
\newblock \doi{10.1103/PhysRevB.95.155129}.

\bibitem[\ifmmode~\check{S}\else \v{S}\fi{}untajs
  et~al.(2020)\ifmmode~\check{S}\else \v{S}\fi{}untajs,
  Bon\ifmmode~\check{c}\else \v{c}\fi{}a, Prosen, and
  Vidmar]{PhysRevE.102.062144}
Jan \ifmmode~\check{S}\else \v{S}\fi{}untajs, Janez Bon\ifmmode~\check{c}\else
  \v{c}\fi{}a, Toma\ifmmode \check{z}\else~\v{z}\fi{} Prosen, and Lev Vidmar.
\newblock Quantum chaos challenges many-body localization.
\newblock \emph{Phys. Rev. E}, 102:\penalty0 062144, Dec 2020.
\newblock \doi{10.1103/PhysRevE.102.062144}.

\bibitem[Abanin et~al.(2019)Abanin, Altman, Bloch, and
  Serbyn]{RevModPhys.91.021001}
Dmitry~A. Abanin, Ehud Altman, Immanuel Bloch, and Maksym Serbyn.
\newblock Colloquium: Many-body localization, thermalization, and entanglement.
\newblock \emph{Rev. Mod. Phys.}, 91:\penalty0 021001, May 2019.
\newblock \doi{10.1103/RevModPhys.91.021001}.

\bibitem[{Alet} and {Laflorencie}(2018)]{2018CRPhy..19..498A}
Fabien {Alet} and Nicolas {Laflorencie}.
\newblock {Many-body localization: An introduction and selected topics}.
\newblock \emph{Comptes Rendus Physique}, 19\penalty0 (6):\penalty0 498--525,
  September 2018.
\newblock \doi{10.1016/j.crhy.2018.03.003}.

\bibitem[{Altman} and {Vosk}(2015)]{2015ARCMP...6..383A}
Ehud {Altman} and Ronen {Vosk}.
\newblock {Universal Dynamics and Renormalization in Many-Body-Localized
  Systems}.
\newblock \emph{Annual Review of Condensed Matter Physics}, 6:\penalty0
  383--409, March 2015.
\newblock \doi{10.1146/annurev-conmatphys-031214-014701}.

\bibitem[{Nandkishore} and {Huse}(2015)]{2015ARCMP...6...15N}
Rahul {Nandkishore} and David~A. {Huse}.
\newblock {Many-Body Localization and Thermalization in Quantum Statistical
  Mechanics}.
\newblock \emph{Annual Review of Condensed Matter Physics}, 6:\penalty0 15--38,
  March 2015.
\newblock \doi{10.1146/annurev-conmatphys-031214-014726}.

\bibitem[Serbyn et~al.(2013)Serbyn, Papi\ifmmode~\acute{c}\else \'{c}\fi{}, and
  Abanin]{PhysRevLett.111.127201}
Maksym Serbyn, Z.~Papi\ifmmode~\acute{c}\else \'{c}\fi{}, and Dmitry~A. Abanin.
\newblock Local conservation laws and the structure of the many-body localized
  states.
\newblock \emph{Phys. Rev. Lett.}, 111:\penalty0 127201, Sep 2013.
\newblock \doi{10.1103/PhysRevLett.111.127201}.

\bibitem[Huse et~al.(2014)Huse, Nandkishore, and Oganesyan]{PhysRevB.90.174202}
David~A. Huse, Rahul Nandkishore, and Vadim Oganesyan.
\newblock Phenomenology of fully many-body-localized systems.
\newblock \emph{Phys. Rev. B}, 90:\penalty0 174202, Nov 2014.
\newblock \doi{10.1103/PhysRevB.90.174202}.

\bibitem[Vosk and Altman(2013)]{PhysRevLett.110.067204}
Ronen Vosk and Ehud Altman.
\newblock Many-body localization in one dimension as a dynamical
  renormalization group fixed point.
\newblock \emph{Phys. Rev. Lett.}, 110:\penalty0 067204, Feb 2013.
\newblock \doi{10.1103/PhysRevLett.110.067204}.

\bibitem[Schreiber et~al.(2015)Schreiber, Hodgman, Bordia, L{\"u}schen,
  Fischer, Vosk, Altman, Schneider, and
  Bloch]{blochObservationManybodyLocalization2015}
Michael Schreiber, Sean~S. Hodgman, Pranjal Bordia, Henrik~P. L{\"u}schen,
  Mark~H. Fischer, Ronen Vosk, Ehud Altman, Ulrich Schneider, and Immanuel
  Bloch.
\newblock Observation of many-body localization of interacting fermions in a
  quasirandom optical lattice.
\newblock \emph{Science}, 349\penalty0 (6250):\penalty0 842--845, August 2015.
\newblock \doi{10.1126/science.aaa7432}.

\bibitem[{Choi} et~al.(2016){Choi}, {Hild}, {Zeiher}, {Schau{\ss}},
  {Rubio-Abadal}, {Yefsah}, {Khemani}, {Huse}, {Bloch}, and
  {Gross}]{2016Sci...352.1547C}
Jae-yoon {Choi}, Sebastian {Hild}, Johannes {Zeiher}, Peter {Schau{\ss}},
  Antonio {Rubio-Abadal}, Tarik {Yefsah}, Vedika {Khemani}, David~A. {Huse},
  Immanuel {Bloch}, and Christian {Gross}.
\newblock {Exploring the many-body localization transition in two dimensions}.
\newblock \emph{Science}, 352\penalty0 (6293):\penalty0 1547--1552, June 2016.
\newblock \doi{10.1126/science.aaf8834}.

\bibitem[{Smith} et~al.(2016){Smith}, {Lee}, {Richerme}, {Neyenhuis}, {Hess},
  {Hauke}, {Heyl}, {Huse}, and {Monroe}]{2016NatPh..12..907S}
J.~{Smith}, A.~{Lee}, P.~{Richerme}, B.~{Neyenhuis}, P.~W. {Hess}, P.~{Hauke},
  M.~{Heyl}, D.~A. {Huse}, and C.~{Monroe}.
\newblock {Many-body localization in a quantum simulator with programmable
  random disorder}.
\newblock \emph{Nature Physics}, 12\penalty0 (10):\penalty0 907--911, October
  2016.
\newblock \doi{10.1038/nphys3783}.

\bibitem[Huang et~al.(2017)Huang, Zhang, and Chen]{Huang:2016knw}
Yichen Huang, Yong-Liang Zhang, and Xie Chen.
\newblock {Out-of-time-ordered correlators in many-body localized systems}.
\newblock \emph{Annalen Phys.}, 529\penalty0 (7):\penalty0 1600318, 2017.
\newblock \doi{10.1002/andp.201600318}.

\bibitem[Fan et~al.(2017)Fan, Zhang, Shen, and Zhai]{Fan:2016ean}
Ruihua Fan, Pengfei Zhang, Huitao Shen, and Hui Zhai.
\newblock {Out-of-Time-Order Correlation for Many-Body Localization}.
\newblock \emph{Sci. Bull.}, 62:\penalty0 707--711, 2017.
\newblock \doi{10.1016/j.scib.2017.04.011}.

\bibitem[{Swingle} and {Chowdhury}(2017)]{2017PhRvB..95f0201S}
Brian {Swingle} and Debanjan {Chowdhury}.
\newblock {Slow scrambling in disordered quantum systems}.
\newblock \emph{Phys. Rev. B}, 95\penalty0 (6):\penalty0 060201, February 2017.
\newblock \doi{10.1103/PhysRevB.95.060201}.

\bibitem[{He} and {Lu}(2017)]{2017PhRvB..95e4201H}
Rong-Qiang {He} and Zhong-Yi {Lu}.
\newblock {Characterizing many-body localization by out-of-time-ordered
  correlation}.
\newblock \emph{Phys. Rev. B}, 95\penalty0 (5):\penalty0 054201, February 2017.
\newblock \doi{10.1103/PhysRevB.95.054201}.

\bibitem[{Chen}(2016)]{2016arXiv160802765C}
Yu~{Chen}.
\newblock {Universal Logarithmic Scrambling in Many Body Localization}.
\newblock \emph{arXiv e-prints}, art. arXiv:1608.02765, August 2016.
\newblock \doi{10.48550/arXiv.1608.02765}.

\bibitem[{Chen} et~al.(2017){Chen}, {Zhou}, {Huse}, and
  {Fradkin}]{2017AnP...52900332C}
Xiao {Chen}, Tianci {Zhou}, David~A. {Huse}, and Eduardo {Fradkin}.
\newblock {Out-of-time-order correlations in many-body localized and thermal
  phases}.
\newblock \emph{Annalen der Physik}, 529\penalty0 (7):\penalty0 1600332, July
  2017.
\newblock \doi{10.1002/andp.201600332}.

\bibitem[Liu et~al.(2024)Liu, Hao, and
  Hu]{liu2024predictingarbitrarystateproperties}
Zhenhuan Liu, Zihan Hao, and Hong-Ye Hu.
\newblock Predicting arbitrary state properties from single hamiltonian quench
  dynamics.
\newblock 2024.
\newblock \doi{10.48550/arXiv.2311.00695}.

\bibitem[{You} et~al.(2018){You}, {Yang}, and {Qi}]{2018PhRvB..97d5153Y}
Yi-Zhuang {You}, Zhao {Yang}, and Xiao-Liang {Qi}.
\newblock {Machine learning spatial geometry from entanglement features}.
\newblock \emph{Phys. Rev. B}, 97\penalty0 (4):\penalty0 045153, February 2018.
\newblock \doi{10.1103/PhysRevB.97.045153}.

\bibitem[{You} and {Gu}(2018)]{2018PhRvB..98a4309Y}
Yi-Zhuang {You} and Yingfei {Gu}.
\newblock {Entanglement features of random Hamiltonian dynamics}.
\newblock \emph{Phys. Rev. B}, 98\penalty0 (1):\penalty0 014309, July 2018.
\newblock \doi{10.1103/PhysRevB.98.014309}.

\bibitem[Fishman et~al.(2022)Fishman, White, and
  Stoudenmire]{10.21468/SciPostPhysCodeb.4}
Matthew Fishman, Steven~R. White, and E.~Miles Stoudenmire.
\newblock {The ITensor Software Library for Tensor Network Calculations}.
\newblock \emph{SciPost Phys. Codebases}, page~4, 2022.
\newblock \doi{10.21468/SciPostPhysCodeb.4}.

\bibitem[Zhou(2024)]{tgzhou_code}
Tian-Gang Zhou.
\newblock {MBL}-{Classical}-{Shadow}.
\newblock \url{https://github.com/tgzhou98/MBL-Classical-Shadow}, 2024.

\bibitem[Qi and Streicher(2019)]{qiQuantumEpidemiologyOperator2019}
Xiao-Liang Qi and Alexandre Streicher.
\newblock Quantum epidemiology: Operator growth, thermal effects, and {{SYK}}.
\newblock \emph{J. High Energ. Phys.}, 2019\penalty0 (8):\penalty0 12, August
  2019.
\newblock ISSN 1029-8479.
\newblock \doi{10.1007/JHEP08(2019)012}.

\bibitem[Yoshida and Kitaev(2017)]{yoshida2017efficient}
Beni Yoshida and Alexei Kitaev.
\newblock Efficient decoding for the hayden-preskill protocol.
\newblock \emph{arXiv e-prints}, art. arXiv:1710.03363, February 2017.
\newblock \doi{10.48550/arXiv.1710.03363}.

\bibitem[Nahum et~al.(2018)Nahum, Vijay, and
  Haah]{nahumOperatorSpreadingRandom2018}
Adam Nahum, Sagar Vijay, and Jeongwan Haah.
\newblock Operator {{Spreading}} in {{Random Unitary Circuits}}.
\newblock \emph{Phys. Rev. X}, 8\penalty0 (2):\penalty0 021014, April 2018.
\newblock \doi{10.1103/PhysRevX.8.021014}.

\bibitem[Kj\"all et~al.(2014)Kj\"all, Bardarson, and
  Pollmann]{pollmannManyBodyLocalizationDisordered2014}
Jonas~A. Kj\"all, Jens~H. Bardarson, and Frank Pollmann.
\newblock Many-body localization in a disordered quantum ising chain.
\newblock \emph{Phys. Rev. Lett.}, 113:\penalty0 107204, Sep 2014.
\newblock \doi{10.1103/PhysRevLett.113.107204}.

\bibitem[Smith et~al.(2016)Smith, Lee, Richerme, Neyenhuis, Hess, Hauke, Heyl,
  Huse, and Monroe]{monroeManybodyLocalizationQuantum2016a}
J.~Smith, A.~Lee, P.~Richerme, B.~Neyenhuis, P.~W. Hess, P.~Hauke, M.~Heyl,
  D.~A. Huse, and C.~Monroe.
\newblock Many-body localization in a quantum simulator with programmable
  random disorder.
\newblock \emph{Nature Phys}, 12\penalty0 (10):\penalty0 907--911, October
  2016.
\newblock ISSN 1745-2481.
\newblock \doi{10.1038/nphys3783}.

\bibitem[Saffman(2016)]{saffmanQuantumComputingAtomic2016}
M.~Saffman.
\newblock Quantum computing with atomic qubits and {{Rydberg}} interactions:
  {{Progress}} and challenges.
\newblock \emph{J. Phys. B: At. Mol. Opt. Phys.}, 49\penalty0 (20):\penalty0
  202001, October 2016.
\newblock ISSN 0953-4075, 1361-6455.
\newblock \doi{10.1088/0953-4075/49/20/202001}.

\bibitem[Monroe et~al.(2021)Monroe, Campbell, Duan, Gong, Gorshkov, Hess,
  Islam, Kim, Linke, Pagano, Richerme, Senko, and
  Yao]{yaoProgrammableQuantumSimulations2021}
C.~Monroe, W.~C. Campbell, L.-M. Duan, Z.-X. Gong, A.~V. Gorshkov, P.~W. Hess,
  R.~Islam, K.~Kim, N.~M. Linke, G.~Pagano, P.~Richerme, C.~Senko, and N.~Y.
  Yao.
\newblock Programmable quantum simulations of spin systems with trapped ions.
\newblock \emph{Rev. Mod. Phys.}, 93\penalty0 (2):\penalty0 025001, April 2021.
\newblock \doi{10.1103/RevModPhys.93.025001}.

\bibitem[Altman et~al.(2003)Altman, Hofstetter, Demler, and
  Lukin]{lukinPhaseDiagramTwocomponent2003}
Ehud Altman, Walter Hofstetter, Eugene Demler, and Mikhail~D. Lukin.
\newblock Phase diagram of two-component bosons on an optical lattice.
\newblock \emph{New J. Phys.}, 5\penalty0 (1):\penalty0 113, September 2003.
\newblock ISSN 1367-2630.
\newblock \doi{10.1088/1367-2630/5/1/113}.

\bibitem[Chin et~al.(2010)Chin, Grimm, Julienne, and
  Tiesinga]{tiesingaFeshbachResonancesUltracold2010}
Cheng Chin, Rudolf Grimm, Paul Julienne, and Eite Tiesinga.
\newblock Feshbach resonances in ultracold gases.
\newblock \emph{Rev. Mod. Phys.}, 82\penalty0 (2):\penalty0 1225--1286, April
  2010.
\newblock \doi{10.1103/RevModPhys.82.1225}.

\end{thebibliography}

\onecolumn\newpage
\appendix

\section{The definition of operator states}\label{app:op_state}
The definition of operator state requires the introduction of the EPR state $|\text{EPR}\rangle = \frac{1}{\sqrt{D}} \sum_{n=1}^{D} |n\rangle_l |n\rangle_r^*$, where $n$ is an arbitrary state in $D$-dimensional basis and $D=2^N$ is the Hilbert space dimension\cite{qiQuantumEpidemiologyOperator2019,yoshida2017efficient}. The symbol ${}^*$ denotes time reversal and $l$ and $r$ denote the left and right system. The operator state is defined by
\begin{equation}
    |O \rangle\rangle = O_l |\text{EPR} \rangle.
\end{equation}
The $l$ subscript in the operator means it's defined in the left system.

The Eq.~\ref{eq:lambda2} can be verified by
\begin{equation}
    \begin{split}
        &D\left(\langle\langle O|U^\dagger\otimes U^T |0\rangle\rangle\right)^2 \\
        =&D\left(\langle \text{EPR}| O_l U^\dagger_l  U^T_r |0\rangle\rangle\right)^2 \\
        =&D\left(\langle \text{EPR}| U_l O_l U^\dagger_l   |0\rangle\rangle\right)^2 \\
        =&D\left(\frac{1}{\sqrt{D}}\langle\langle 0| U_l O_l U^\dagger_l   |0\rangle\rangle\right)^2 \\
        =&\langle 0|U OU^\dagger|0\rangle^2 \\
    \end{split}
\end{equation}
The third line has used the property of the EPR state $\langle \text{EPR} | U_r^T = \langle \text{EPR} | U_l$ \cite{qiQuantumEpidemiologyOperator2019,yoshida2017efficient}. The fourth line is due to only the $\langle\langle 0 |$ term in the EPR state will lead to the non-zero overlap.

\section{Average of local Haar random unitary gates}\label{app:haar_average}
The average of $\otimes_j u_j$ can be done first, as discussed in the literature \cite{PhysRevLett.130.230403}. It leads to the $|\mathcal{S}))$ state defined by Eq.~\eqref{eq:S_op2_state}. It results in the following formula
\begin{equation}
    \lambda_O=\mathds{E}_v \langle\langle [\otimes_j v_j^{\dagger}] O_A [\otimes_j v_j] |^{\otimes 2}(e^{iHt}\otimes e^{-iH^Tt} )^{\otimes 2}|\mathcal{S})) 
\end{equation}
To evaluate the $\otimes_j v_j$, we can focus on the left bra state. The operator can be written as the operator in the left system which applies to the EPR state 
\begin{equation}
    \begin{split}
        &\mathds{E}_v \langle\langle [\otimes_j v_j^{\dagger}] O_A [\otimes_j v_j] |^{\otimes 2} \\
        =& \mathds{E}_v \langle \text{EPR}|_1 \langle \text{EPR}|_2 [\otimes_j v_{j,l1}^{\dagger}] O_{A,l1} [\otimes_j v_{j,l1}] [\otimes_j v_{j,l2}^{\dagger}] O_{A,l2} [\otimes_j v_{j,l2}]  \\
    \end{split}
\end{equation}
Here the $l,r$ index corresponds to operator space, $1,2$ index stands for the extra double space. 

Recalling the formula for Haar random unitary average \cite{nahumOperatorSpreadingRandom2018}, we focus on the single site result and omit the site index $i$ on $v$ and assume region $A$ only has one site temporarily. The average of the four copies of $v$ reads
\begin{equation}
    \begin{split}
       Q_A  =&\mathds{E}_v  \left( v_1|a\rangle (O_{A,l1})_{ab} \langle b| v_1^{\dagger} \right) \otimes \left( v_2|c\rangle (O_{A,l2})_{cd} \langle d| v_2^{\dagger} \right) \\
        =&(O_{A,l1})_{ab} (O_{A,l2})_{cd} \left( \sum_{s=\pm} \frac{\overline{\mathds{I}}_{l} + s \overline{F_{l}}}{2q(q+s)} \left( \delta_{ab}\delta_{cd} + s \delta_{cb}\delta_{ad} \right) \otimes \overline{\mathds{I}}_{r} \right) \\
    \end{split}
\end{equation}
Here the overlines in $\overline{\mathds{I}}_{l}, \overline{F_{l}}$ mean the operator is spanned on double space $1,2$ space. $\overline{\mathds{I}}_{l}$ means trivial operation on the double space and $\overline{F_{l}}$ means swap the $1,2$ space. $q$ is the local Hilbert space dimension. For convenience, the state is the eigenstate of $Z$ operators and indexes $a,b,c,d$ are implicitly summed up.

We separately discuss the result. If $O_A$ is non-trivial on the current site, i.g. $O_A = X,Y,Z$ on this site, the result can be simplified as
\begin{equation}
    Q_A = \frac{\overline{\mathds{I}}_{l} - q \overline{F_{l}}}{1-q^2}  \otimes \overline{\mathds{I}}_{r}
\end{equation}
If $O_A$ is not supported on the site, i.g. $O_A = \mathds{I}$, then we have
\begin{equation}
    Q_A = \overline{\mathds{I}}_{l}  \otimes \overline{\mathds{I}}_{r}
\end{equation}

Finally, we apply $Q_A$ on the double EPR state, and use the fact that the SWAP operator can be written in the following form $\overline{F}_l = \frac{1}{q} \sum_{P} P_{l1} P_{l2} $, then the state for each site in $((A |$ state can be represented by

\begin{equation}
((\psi_{i} |= \begin{cases}
       \langle \text{EPR}|_1 \langle \text{EPR}|_2 \frac{ \sum_{P\neq I} P_{l1} P_{l2}}{q^2-1} \otimes \overline{\mathds{I}}_{r} & \text{for $i\in A$,}\\
      \langle \text{EPR}|_1 \langle \text{EPR}|_2 \overline{\mathds{I}}_{l} \otimes \overline{\mathds{I}}_{r} & \text{for $i\notin A$.}
    \end{cases}   
\end{equation}
Finally, we take $q=2$ and obtain the Eq.~\eqref{eq:A_op2_state} in the main text.

\section{Phenomelogical model}
\subsection{Proof of Statement 1}\label{supp_ssub:Proof_Theorem1}
\begin{proof}
    According to the definition of the unitary $U$ in our protocol (Eq.~\eqref{eq:unitary_Shadow}), the right-side local Haar unitary gate transforms any local operator $O_i$ into an equal mixture of Pauli operators $\tilde{X}_i$, $\tilde{Y}_i$, and $\tilde{Z}_i$. As a result, the initial operators $\tilde{Z}_i$ appear with probability $P_0(\tilde{Z}_i) = 1/3$, while $\tilde{X}_i$ or $\tilde{Y}_i$ appear with probability $P_0(\tilde{X}_i, \tilde{Y}_i) = 2/3$. Following the discussions in Section~\ref{sec:PhenomelogicalModel}, we know that the initial operator $\tilde{Z}_i$ contributes a factor of $2/3$, and $\tilde{X}_i$ or $\tilde{Y}_i$ contribute a factor of $1/3$ to $\overline{\sum_{n_i} P(n_i)3^{-n_i}}$. Therefore, $\lambda_O$ for sites in the subsystem $A$ with length $k$ is given by:
\begin{equation}\label{eq:lambda_in_A}
     \prod_{i \in A} \overline{\sum_{n_i} P(n_i)3^{-n_i}} 
     = \left(P_0(\tilde{Z}_i) \frac{2}{3} + P_0(\tilde{X}_i,\tilde{Y}_i) \frac{1}{3} \right)^k
     = \left(\frac{9}{4} \right)^k 
\end{equation} 
 We also need to include the contributions from outside the subsystem $A$ but within the logarithmic light cone, which reads
\begin{equation}\label{eq:lambda_notin_A}
    \prod_{i \notin A} \overline{\sum_{n_i} P(n_i)3^{-n_i}} 
     = \left(\frac{2}{3} \right)^{\Delta N} \\
     = \left(\frac{2}{3}\right)^{2\xi \log (J_0 t)}.
\end{equation}
Combing Eq.~\eqref{eq:lambda_in_A},~\eqref{eq:lambda_notin_A}, it finally leads to Eq.~\eqref{eq:reslambdaeff}.
\end{proof}

\subsection{High order term contributions}\label{supp_ssub:High_order}
We can address this from two perspectives. First, by combining second-order and higher-order terms, we obtain an effective two-body interaction. As argued in Eq.~(2) of the literature \cite{PhysRevB.90.174202}, the effective two-body interaction $J_{ij}^{\text{eff}} = J_{ij} + \sum_n J^{(n)}_{i\{k\}j} \tilde{Z}_{k_1} \tilde{Z}_{k_2} \cdots \tilde{Z}_{k_n}$ still decays with distance $r$, i.e., $J^{\text{eff}} \sim J_0 \exp(-r/\xi)$. This results in the same operator dynamics as when only considering the two-body pseudospin interaction.

Secondly, we can argue that the updating rule for determining $\lambda_O$ (and therefore the shadow norm) remains unchanged after including higher-order interactions. Physically, this is because higher-order terms also preserve the LIOMs, or more specifically, $\tilde{Z}_k$. For our example with an initial operator $\tilde{Z}_1 \tilde{X}_2$, higher-order interactions, such as $\sum_{ijk} J_{ijk} \tilde{Z}_{i} \tilde{Z}_{j} \tilde{Z}_{k}$, still mix $\tilde{Z}_1$ and $\tilde{I}_1$ on site 1, and mix $\tilde{X}_2$ and $\tilde{Y}_2$ on site 2. Other sites can also be occupied by operators $\tilde{Z}_{i}$ and $\tilde{I}_{i}$. In the long-time limit, we know that sites occupied by $\{\tilde{Z}, \tilde{I}\}$ will contribute $\frac{1}{2} \times \frac{1}{3} + \frac{1}{2} = \frac{2}{3}$ to $\lambda_O$. Similarly, sites occupied by $\{\tilde{X}, \tilde{Y}\}$ will contribute $\frac{1}{3}$ to $\lambda_O$, as all possible choices are non-trivial operators. This leads to the same result as when considering only two-body interactions.

\section{Experimental Realization of MBL-Based Classical Shadow}\label{app:exp_proposal}
In this work, our main argument is that analog quantum evolution for the MBL Hamiltonian is a type of Hamiltonian-driven algorithm that realizes the efficient classical shadow algorithm. This approach differs from the digital realization of global Haar evolution using two-qubit gates
\cite{PhysRevResearch.4.013054,PhysRevResearch.5.023027,2023Quant...7.1026A,jaffeClassicalShadowsPauliinvariant2024,PhysRevLett.130.230403}.

The experimental realization of the MBL-based classical shadow protocol requires the compatible implementation of single-qubit random Clifford gates and the MBL Hamiltonian. We show that this can be achieved using two different types of MBL Hamiltonians and the corresponding quantum platforms.

Experimentally, the first candidate Hamiltonian to realize MBL dynamics can be the disordered quantum Ising model (DQIM)
\begin{equation}
    H_{\text{DQIM}} = \sum_{ij} J_{ij} Z_{i} Z_{j} + \sum_i h_i X_{i} 
\end{equation}
where $J_{ij}$ can include only nearest-neighbor or next-nearest-neighbor interactions and is uniformly sampled from $J_{ij} \sim \text{Uni}[J-\delta J, J+\delta J]$. $h_i$ is the transverse field, which can be randomly sampled or not. The DQIM Hamiltonian can induce non-ergodic MBL dynamics by increasing the value of $\delta J/J$ \cite{pollmannManyBodyLocalizationDisordered2014,monroeManybodyLocalizationQuantum2016a}.

It is known that this Hamiltonian can be well simulated by both Rydberg atom \cite{saffmanQuantumComputingAtomic2016} and trapped ion \cite{yaoProgrammableQuantumSimulations2021} based quantum simulators. However, there is a side remark that the DQIM Hamiltonian implicitly requires atoms in the Rydberg state, while the qubits associated with the digital gate are usually encoded on clock states of the ground state manifold \cite{kasevichAtomicInterferometryUsing1991,riehleOpticalRamseySpectroscopy1991,levineHighFidelityControlEntanglement2018,levineDispersiveOpticalSystems2022,bluvsteinQuantumProcessorBased2022,bluvsteinLogicalQuantumProcessor2024}. The different sets of quantum bases could induce more errors in the conversion process. This inconsistency shows that the DQIM Hamiltonian-driven classical shadow protocol is not suitable for Rydberg neutral atoms. However, it still works for the trapped ion platform, making our MBL algorithm compatible with promising NISQ quantum devices.

Secondly, for the neutral atoms platform, we can realize the random field XXZ model as in Eq.~(3) in the main text by taking the deep Mott region of the interacting bosonic Aubry-Andr\'e model (or fermionic Aubry-Andr\'e model). In principle, the randomness has been realized by two incommensurate lattices to achieve the fermionic Aubry-Andr\'e model, inspired by pioneering experimental work \cite{blochObservationManybodyLocalization2015}. To adapt to the qubit system in the classical shadow, we require two components for the Bose-Hubbard model \cite{lukinPhaseDiagramTwocomponent2003} (or the spin-up and spin-down components for the Fermi-Hubbard model). This can be realized in the optical lattice with a tunneling effect, without requiring excitation to the Rydberg states. The Hamiltonian reads
\begin{equation}
    \begin{split}
        H_{\text{AA}} =& -  \sum_{i,\sigma} t_{\sigma} \left(b^{\dagger}_{i,\sigma} b_{i+1, \sigma} + \text{h.c.}  \right) + \sum_{i,\sigma} \Delta_{\sigma} \cos(2\pi\beta i+ \phi) b^{\dagger}_{i,\sigma} b_{i, \sigma}  \\
        &+ \sum_{i} U (n_{i,1}-1/2) (n_{i,2}-1/2) + \sum_{i,\sigma} V_{\sigma}n_{i,\sigma}(n_{i,\sigma}-1) - \sum_{i,\sigma} \mu_{\sigma} n_{i,\sigma}, \\
    \end{split}
\end{equation}
where $b^{\dagger}_{i,\sigma},b_{i,\sigma}$ are either bosonic or fermionic creation and annihilation operators, and $\sigma=1,2$ labels two components. $n_{i,\sigma}$ is the number operator, and the intra-component interaction term $V_{\sigma}n_{i,\sigma}(n_{i,\sigma}-1)$ vanishes in the case of fermions. $J_{\sigma}$ represents the species-dependent tunneling energy for the optical lattices. $\Delta_{\sigma}$ represents the second quasiperiodic potential, where $\beta$ is close to an irrational number if the wavelengths of two sets of optical lattices are incommensurate. $U$ is the inter-component Hubbard interaction induced by Feshbach resonance \cite{tiesingaFeshbachResonancesUltracold2010}. In the deep Mott limit $J_{\sigma} \ll U, V_\sigma$, we consider a total filling of one particle per site, where the effective Hamiltonian of the bosonic case can be simplified as the random field XXZ model with a random $h$ field and random spin interactions 
\begin{equation*}
     H_{\text{XXZ}} = J \sum_i(X_{i}X_{i+1}+Y_{i}Y_{i+1}+\Delta Z_{i}Z_{i+1}) +\sum_ih_i Z_i,
\end{equation*}
where
\begin{equation}
     \begin{split}
         J &= - \frac{4 t_1 t_2}{U} \\
         J \Delta &= 2 \frac{t_1^2 + t_2^2}{U} - \frac{4 t_1^2}{V_1} - \frac{4 t_2^2}{V_2} \\
         h_i &= - \frac{2 t_1^2}{V_1} + \frac{2 t_2^2}{V_2} - (\mu_1 - \mu_2) + (\Delta_1 - \Delta_2) \cos(2\pi\beta i+\phi)
     \end{split}
\end{equation}
. Here, large $U$, $V_{\sigma}$ excludes the possibility of doublons and ensures that the original qubit system is still well-defined after the analog MBL Hamiltonian evolution.

\end{document}